\documentclass[aps,showkeys,preprint]{revtex4-1}
\usepackage[T1]{fontenc}
\usepackage{amsthm}
\usepackage{amsmath}

\usepackage{amssymb}
\usepackage{graphicx}
\usepackage{color}
\usepackage{natbib}
\usepackage{soul}

\makeatletter

\begin{document}

\title{Phase diagram of the mixed-spin (1,3/2) Ising ferrimagnetic system with two different anisotropies}

\author{D. C. da Silva}\email{davidcristiano@fisica.ufmt.br}
\author{A. S. de Arruda}\email{aarruda@fisica.ufmt.br}
\author{M. Godoy}\email{mgodoy@fisica.ufmt.br}
\affiliation{Instituto de F\'{\i}sica, Universidade Federal de Mato Grosso, 78060-900, Cuiab\'{a}, MT, Brazil.}
\date{\today}
\begin{abstract}
In this work, we have performed Monte Carlo simulations to study phase transitions in a mixed spin-1 and spin-3/2 Ising ferrimagnetic system on the square and cubic lattices and with two different single-ion anisotropies. This lattice is divided in two interpenetrating sublattices with spins $S^A = 1$ (states $\pm1$ and 0) on the sublattice $A$ and $S^B = 3/2$ (states $\pm 3/2$, $\pm 1/2$) on the sublattice $B$. We have used single-ion anisotropies $D_{A}$ and $D_{B}$ acting on the sites of the sublattice $A$ and $B$, receptively. We have determined the phase diagrams of the model in the temperature $T$ versus the single-ion anisotropies strength $D_A$ and $D_B$ plane and shown that the system exhibits both second- and first-order phase transitions. We also have shown that this system displays compensation points for some values of the anisotropies.

\end{abstract}

\maketitle

\section{Introduction}
The study of mixed-spin models has its importance recognized because they are related to ferrimagnetic materials \cite{mm2,mm3,mm4}. In these models, the  particles that carry two different spins are  distributed in two interpenetrating sublattices. Then each pair nearest-neighbor spins are coupled antiferromagnetically so that at low temperatures all the spins are allied antiparallel.  Thus, in each sublattice, there is a magnetization, both with different magnitudes and opposite signs. Therefore, the system as a whole presents a magnetization (total magnetization). When the temperature is increased, the spins of the different sublattices have their alignments decreased. Then, at a certain temperature, these alignments inverted of the magnetic moments are compensated, causing the total magnetization goes to zero for a temperature smaller than the critical temperature ($T_c$). This temperature is called the compensation temperature ($ T_{ comp}$). Materials that present this behavior are known as ferrimagnets.  The compensation temperature becomes the ferrimagnet system of great interest for technological applications \cite{man,hpd,mol} because in this point the coercive field has great growth \cite{hansen,buendia}, so a small driving field is necessary to change the signal of the resulting magnetization.
\par
Most ferrimagnetic materials have been modeled by mixed-spins Ising model through a variety of combinations of two spins $ (\sigma, S) $, i.e., $( 1/2, 1)$, $( 1/2, 3/2)$, $( 1, 3/2)$, and so on.  It is important to note that the critical behavior is the same for both ferromagnetic ($J> 0$) and ferrimagnetic ($J<0$) systems \cite{Abubrig}. There are  exact solutions \cite{lin, lipo, jas, dak} for the simplest case $( 1/2, 1)$. Kaneyoshi {\it et al.} \cite{Kaneyoshi1,Kaneyoshi2} and Plascak {\it et al.} \cite{Resende} provided  theoretical investigations of magnetic properties and the influence of a single-ion anisotropy in the compensation temperature $(T_{comp})$ of a bipartite ferrimagnets such as $MnCu(pba-OH)(H_2)_3$. These spin systems have been  investigated using  a variety of approaches, such as effective-field theory \cite{Kaneyoshi0,Kaneyoshi3,Kaneyoshi4, Benyoussef1,Benyoussef2,Bobak2,ertas1,ertas2}, mean-field approximation \cite{ita1, ze1, ze2, espri1, espri2, Abubrig}, renormalization-group \cite{Salinas}, numerical Monte Carlo simulations \cite{Zhang,Buendia2,godoy}. In particular, the model that presents the combination of spins ($ S_1 = 1 $ and $ S_2 = 3/2 $) has been too much studied, so that there is synthesized material $[NiCr_2 (bipy) _2 (C_2O_4)_4 (H_2 O)] H_2 O $, which indicates  a very rare case of the existence of antiferromagnetism between $ Ni$ with $ S = 1 $ and $Cr$ with $ S = 3/2 $ \cite{Stanica}.  On the other hand, from the theoretical point of view, Abubrig {\it et al.} \cite{ Abubrig} and Souza {\it et al.} \cite{ita1} performed mean-field studies and showed that the complete phase diagram exhibited a tricritical behavior and compensation points.
 \par
In this paper, we are interested to study the phase diagram, giving a greater emphasis on the first-order phase transitions.  We have also looked for occurrences of compensation temperatures by using  Monte Carlo simulations.  Thus, we have inspired in the work of Pereira {\it et al.} \cite{valim} who performed Monte Carlo simulations to study a mixed spin-1 and spin-3/2 Ising ferrimagnetic system on a square lattice with two different random single-ion anisotropies. They have determined the phase diagram of the model in the  temperature  versus strength random single-ion anisotropy plane showing that it exhibits only second-order phase transition lines, and they also have shown that this system displays compensation temperatures for some cases of the random single-ion distribution. Here, using a case more complete we have shown that the system also presents first-order phase transition and tricritical behavior.
\par 
The paper is organized as follows: in Section II, we have described the mixed spin-1 and spin-3/2  ferrimagnetic system and we present some details concerning the simulation procedures. In Section III, we have shown the results obtained. Finally, in the last Section IV, we have presented our conclusions.

\section{The model and simulations}

The mixed spin-1 and spin-3/2 Ising ferrimagnetic system consists of two interpenetrating square and cubic sublattices $A$, with spin-1 (states $S^{A}=0, \pm 1$), and $B$ with spin-3/2 (states $S^{B}= \pm 1/2, \pm 3/2$). In each site of the sublattices, there are single-ion anisotropies $D_A$ and $D_B$ acting on the spin-1 and spin-3/2, respectively. This system is described by the following Hamiltonian model,
\begin{equation}
{\mathcal H}=-J\sum_{\left<i,j\right>}S_{i}^{A}S_{j}^{B} + D_A \sum_{i \in A}(S_{i}^{A})^{2} + D_B \sum_{j \in B} (S_{j}^{B})^{2}, \label{eq1}
\end{equation}
where the first term represents the interaction between the nearest neighbors spins on sites $i$ and $j$ located on the sublattices $A$ and $B$, respectively. $J$ is the magnitude of the exchange interaction, and the sum is over all nearest neighboring pairs of spins. 
$ J$ may be either antiferromagnetic, $J < 0$, as assumed often for ferrimagnets, or ferromagnetic, $J > 0$. Both cases are completely equivalent by a simple spin reversal on either sublattice. Here, for the reason of simplicity, we have considered the case ferromagnetic exchange interaction, $J>0$ in our simulations. As a consequence, in our case, the magnetizations of both sublattices are identical at the compensation point, while in the antiferromagnetic case, at the same compensation point, the sublattice magnetizations have equal magnitude but different sign leading to the above-mentioned vanishing the total magnetization before the critical temperature \citep{godoy2}. The second  and third terms represent the single-ion anisotropies $D_A$ and $D_B$ at all the sites of the sublattices $A$ and $B$, respectively. Therefore, the sum is only performed over $N/2$ spins of the sublattices $A$ and $B$. 

The magnetic properties of the system have been studied using Monte Carlo simulations. In our simulations were used lattice sizes ranging from $L=16$ up to 128 for the square lattice and from $L=8$ up to 32 for the cubic lattice. These lattices consist of two interpenetrating sublattices, each one containing $L^2/2$ (square lattice) and  $L^3/2$ (cubic lattice) sites with periodic boundary conditions. The initial states of the system were prepared randomly and updated by the Metropolis algorithm \cite{metro}. We used $10$ independent samples for any size lattice, but as the error bars are smaller than the symbol sizes, we do not show it in the figures. Typically, we used $3.0 \times 10^5$ MCs (Monte Carlo steps) for the calculation of average values of thermodynamic quantities of interest, after discarding $1.0\times 10^5$ MCs for thermalization, for both square and cubic system. Here, 1 MCs means $N=L^2$ (square lattice) or $L^3$ (cubic lattice) trials to change the state of a spin of the lattice.  The temperature is measured in units $J/k_B$ (equal 1.0 for all simulations) and the anisotropies are measured in units $J/zk_B$, where $z$ is the coordination number and $z=4$ for the square and $z=6$ for cubic lattices.

We have calculated the sublattice magnetizations per site, $m_A$ and $m_B$, defined as
\begin{equation}
m_A = \frac{2[\langle M_A\rangle]}{N} = \frac{2\left[\left<\sum_{A} S_{i}^A \right>\right]}{N},
\end{equation}
and
\begin{equation}
m_B = \frac{2[\langle M_B\rangle]}{N} = \frac{2\left[\left<\sum_{B} S_{j}^B \right>\right]}{N},
\end{equation}
where $\langle \cdots \rangle$ denotes thermal averages and $[\cdots ]$ denotes average over the samples of the system. The order parameter is the total magnetization per site $m_T$ defined as
\begin{equation}
m_T = \frac{[\langle M\rangle]}{N} = \frac{[\langle M_A + M_B\rangle]} {N} = \frac{|m_A + m_B|}{2}.
\end{equation}
Therefore, we defined another parameter that is convenient to obtain the compensation point, which is given by
\begin{equation}
m_S = \frac{[\langle M\rangle]}{N} = \frac{[\langle M_A - M_B\rangle]}{N} = \frac{|m_A - m_B|}{2}.
\end{equation}
Further, we also have calculated the following thermodynamics quantities, the specific heat per site
\begin{equation}
 C_e = \frac{[\langle E^2\rangle]  - [\langle E\rangle]^2}{k_B T^2 N},
\end{equation}
where $k_B$ is the Boltzmann constant and $E$ the total energy of the system. The susceptibility is denoted by $\chi$:
\begin{equation}
\chi = \frac{[\langle M_T^2\rangle]  -  [\langle M_T\rangle] ^2}{k_B T N},
\end{equation}

In order to find the critical point, we used the total $U$ fourth-order Binder cumulant \cite{bin} defined by:
\begin{equation}
U = 1 - \frac{[\langle M_T^4\rangle] }{3[\langle M_T^2\rangle]^2},
\end{equation}

The transition temperature also can be estimated by the position of the peaks of the response functions $C_e$ and $\chi$, but to obtain with greater accuracy in some cases, we have used the intersection of the curves of fourth-order Binder cumulants for different lattice sizes $L$.

The parameter $m_S$ vanishes at the compensation point \cite{so}. Then, the compensation point can be determined by looking at the point where the sublattice magnetizations would coincide. We also require that o compensation point occurs for temperatures below $T_c$, where $T_c$ is the critical temperature.

\section{Results and discussions}

\subsection{Ground-state}

We begin by presenting the ground-state diagram of the system. The ground-state is similar to the obtained by Abubrig et al. and Nakamura \cite{Abubrig, naka}, but in our case we have used in the Hamiltonian of the system the sign plus for $D_A$ and $D_B$ (see Eq. \ref{eq1}) and the exchange parameter is $J>0$. At zero temperature, we have found four phases with different values of $\{m_A, m_B, q_A, q_B\}$, namely the ordered ferrimagnetic phases $F_I \equiv \{1, 3/2,1,9/4 \}$ (or $F_I \equiv \{-1, -3/2,1,9/4 \}$), $F_{II} \equiv \{1,1/2,1,1/4 \}$ (or $F_{II} \equiv \{-1,-1/2,1,1/4 \}$) and the disordered phases $P_I \equiv \{0,0,0,9/4 \}$ and $P_2 \equiv \{0,0,0,1/4 \}$, where the parameters $q_A$ and $q_B$ are defined by $q_A = \langle(S_i^{A})^2 \rangle$ and $q_B = \langle(S_j^{B})^2 \rangle$. 

\begin{figure}[h]
\centering
\includegraphics[scale=0.7]{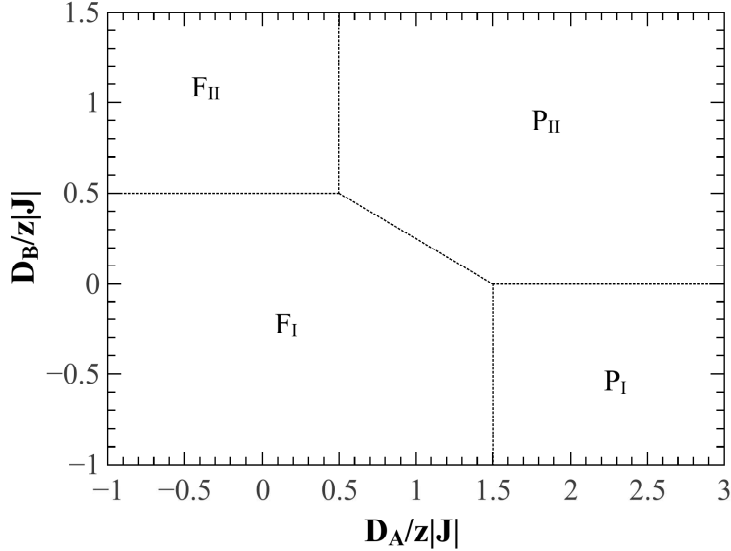} 
\caption{Ground-state diagram of the mixed spin-1 and spin-3/2 Ising ferrimagnetic system with  two different single-ion anisotropies $D_A/z|J|$ and $D_B/z|J|$. The four phases: ordered $F_{I}$, $F_{II}$ and disordered $P_{I}$, $P_{II}$ are separated by lines of first-order transitions. } 
\label{fig1}
\end{figure}

Therefore, the ground-state phase diagram is easily obtained from Hamiltonian (Eq. \ref{eq1}) by comparing the ground-state energies of the different phases and we have shown in Fig. \ref{fig1}. All the phases are separated by lines of first-order transitions represented by dotted lines and the values of the coordinates are obtained independently of the coordination number $z$. For an interesting case, on the line $D_B/z|J|=0.5$, we have $F_I$ and $F_{II}$ coexisting phases. On the other hand, for very small temperatures (for example $T \approx 0.1$) the sublattice magnetizations are $m_A \cong 1.0$ and $ m_B \cong 1.5$ (see  Fig. \ref{fig8}(a) and (b)) and when the temperature increases they go together continuously to zero. Thus, we will call this phase of the ordered phase $F$.

\subsection{Phase diagram for the case $D=D_A/J=D_B/J$}

\begin{figure}[h]
\centering
\includegraphics[scale=0.6]{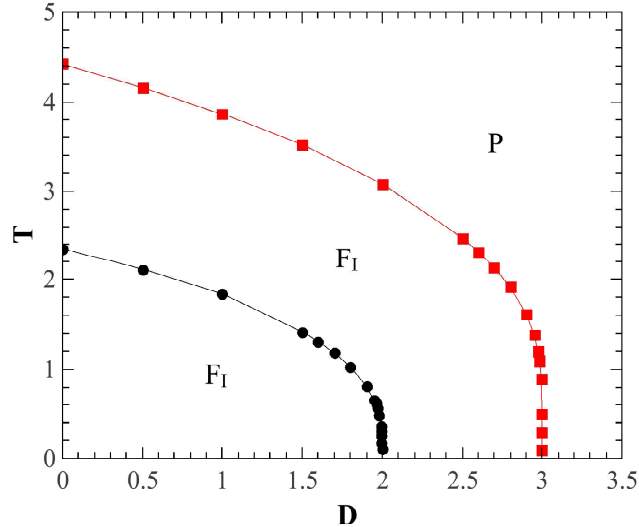}
\caption{(Color online) Phase diagram in the $D-T$ plane for mixed spin-1 and spin-3/2 Ising ferrimagnetic system on the square and cubic lattices. Here, we considered the case $D=D_A=D_B$. The circle- (square lattice) and square-solid (cubic lattice) lines denote second-order phase transitions.}
\label{fig2}
\end{figure}

We also have calculated the phase diagram in the $D-T$ plane for the mixed spin-1 and spin-3/2 Ising ferrimagnetic system on the square and cubic lattice, and for the case $D=D_A/J=D_B/J$, as shown in Fig. \ref{fig2}. Our results obtained on the square lattice are similar to the obtained by \v{Z}ukovi\v{c}  and Bob\'{a}k \cite{milo}, but with the difference that here we have used in the Hamiltonian (see Eq. \ref{eq1}) sign plus for $D=D_A/J=D_B/J$ and the exchange parameter is $J>0$. The phase diagram exhibits only second-order phase transitions between the ordered $F_I$ and disordered $P$ phases on the square (circle-solid line) and cubic (square-solid line) lattices. 

\begin{figure}[h]
\centering
\includegraphics[scale=0.4]{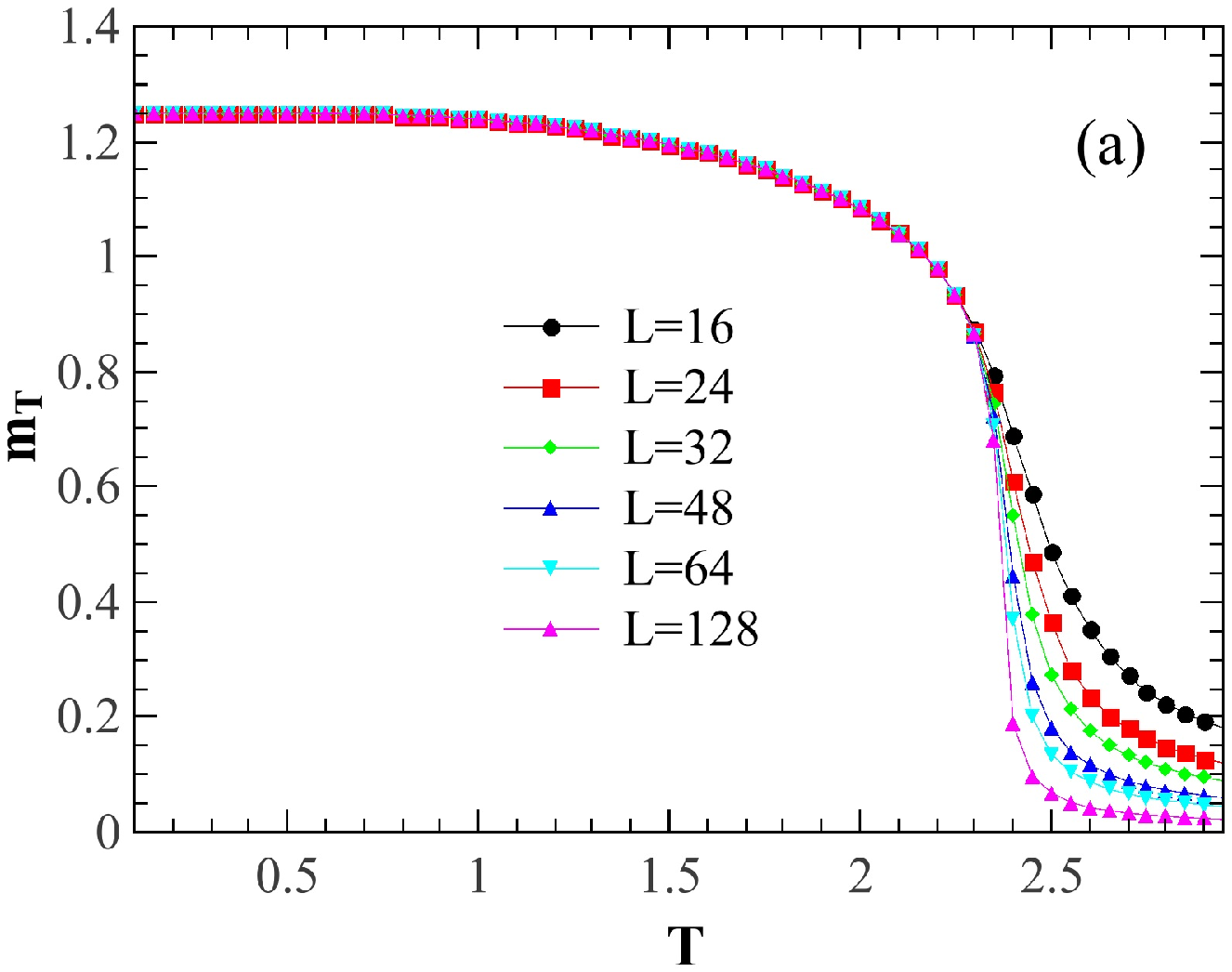} 
\includegraphics[scale=0.4]{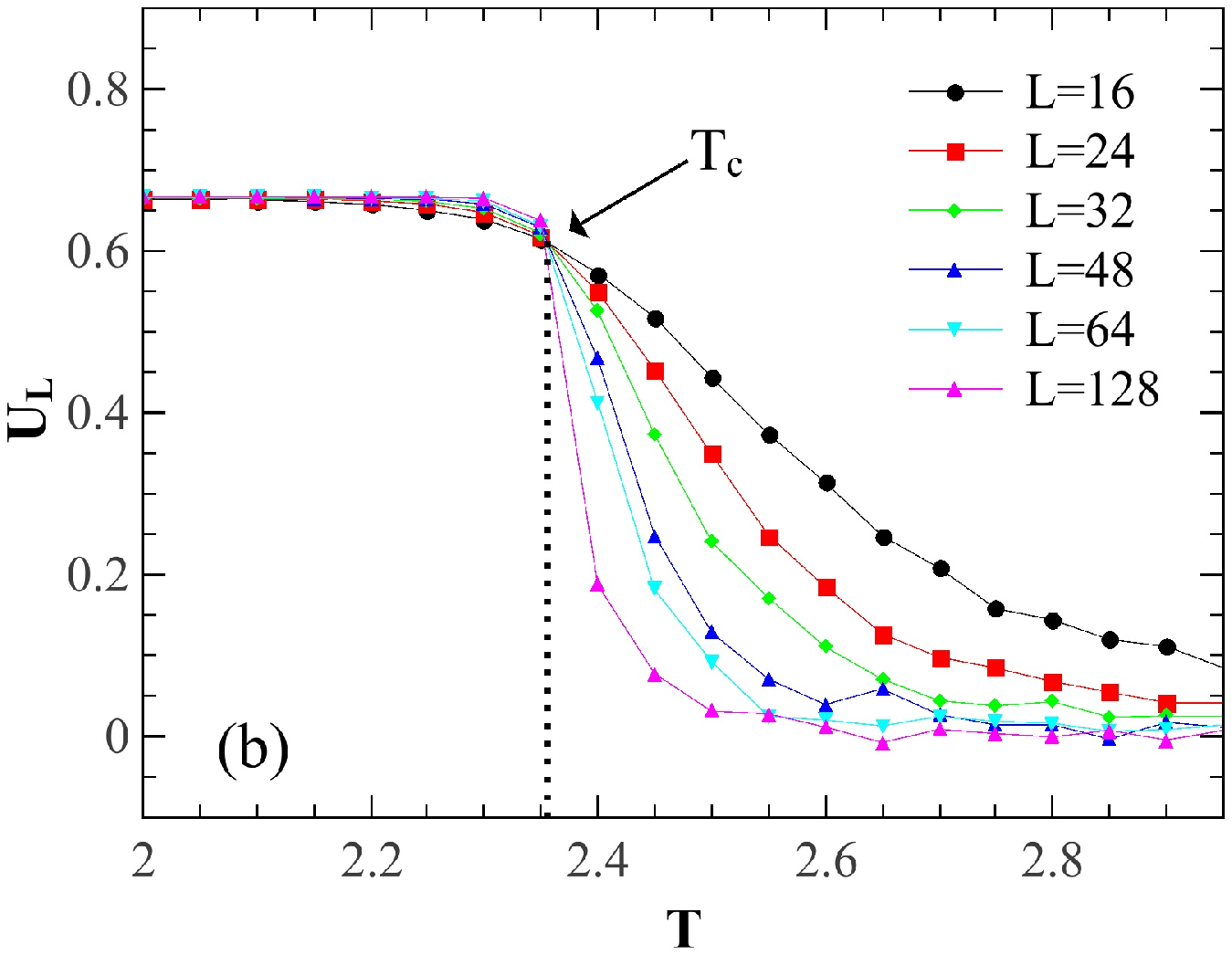}
\includegraphics[scale=0.4]{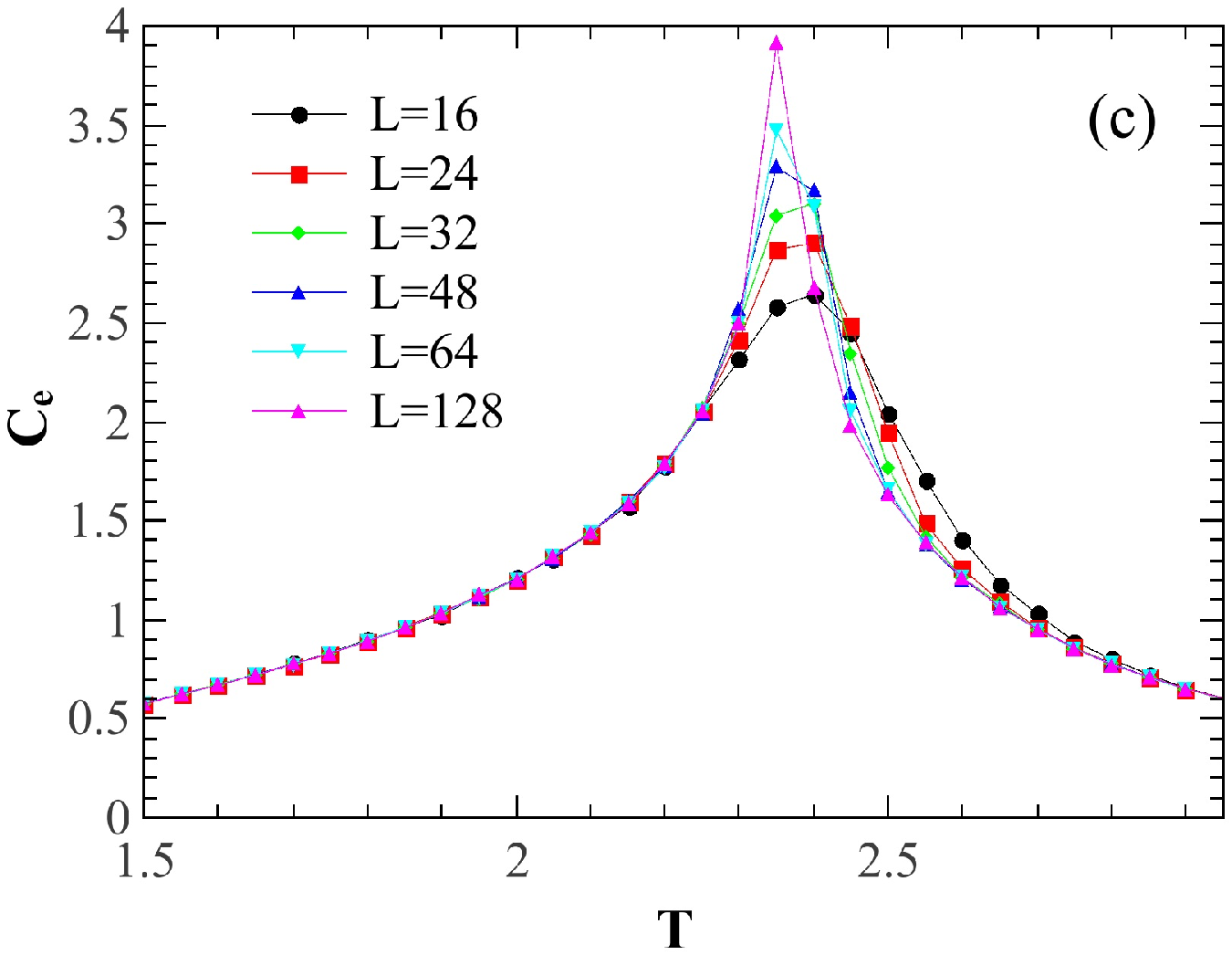}
\includegraphics[scale=0.4]{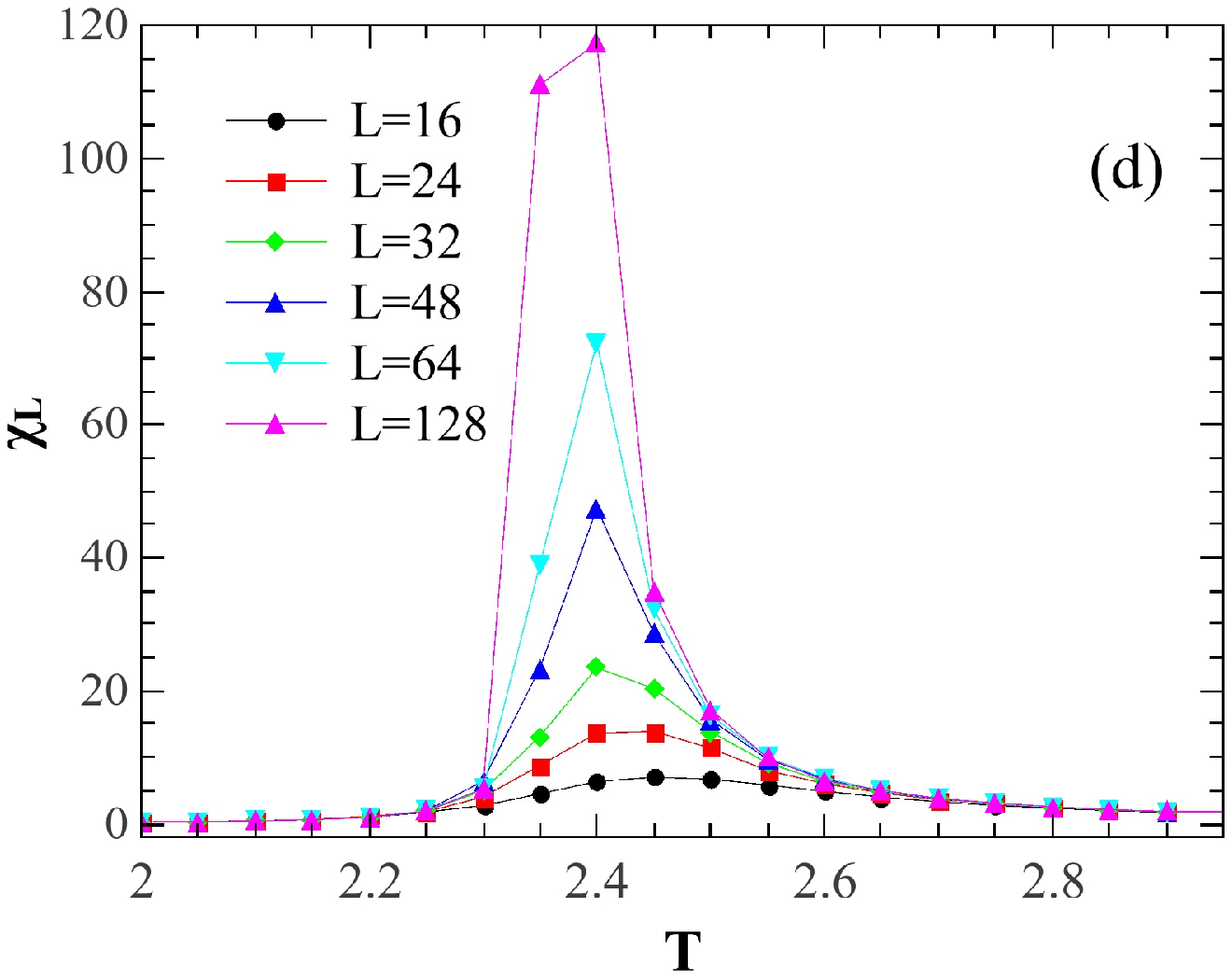}
\caption{(Color online) (a) Total magnetization $m_T$, (b) fourth-order cumulant $U_L$, (c) specific heat $C_e$ and (d) susceptibility $\chi_L$ as a function of temperature $T$ and for different lattice sizes, as indicated in the figures. Here, we considered the case $D=0$.}
\label{fig3}
\end{figure}

In order to observed the finite-size behavior of the magnetic properties of the system, we have calculated the total magnetization $m_T$ (Fig. \ref{fig3}(a)), the fourth-order cumulant $U_L$ (Fig. \ref{fig3}(b)), the specific heat $C_e$ (Fig. \ref{fig3}(c)) and the susceptibility $\chi_L$ (Fig. \ref{fig3}(d)) as a function of temperature $T$ and for different lattice sizes, as indicated in Fig. \ref{fig3}.

The transition points can be estimated from the locations of the peaks of the specific heat $C_e$ and of the susceptibility $\chi_L$. To find the critical points with higher precision, we can use the intersection of the fourth-order cumulant $U_L$ curves for different lattice size (see Fig. \ref{fig3}(b)) \cite{bin}. Thus, we obtained the value of critical temperature with $D=0$, as $T_c=2.354 \pm 0.003$ (square lattice), where this value is in good agreement with one found in the reference \cite{milo}. We also obtained $T_c=4.419 \pm 0.002$ on the cubic lattice with $D=0$. To find the coordinates of the points in the transition lines for high temperatures (see Fig. \ref{fig2} for $1.0 \leq T < T_c (D=0$)), we have used the peaks of the susceptibility as a function of temperature $T$. On the other hand, in the region of low temperatures ($0 < T \lesssim 1.2$) the phase boundary is almost vertical to the $D$-axis, instead of the temperature dependencies of various thermodynamic functions it is more convenient to look into their single-ion parameter $D$ dependencies at a fixed temperature $T$. Therefore, we have used the peaks of the susceptibility $\chi$ as a function of the anisotropy $D$ for a $T$ fixed. In all simulations, we have used  $L=128$ and $L=32$ on the square and cubic lattices, respectively.

\begin{figure}[h]
\centering
\includegraphics[scale=0.4]{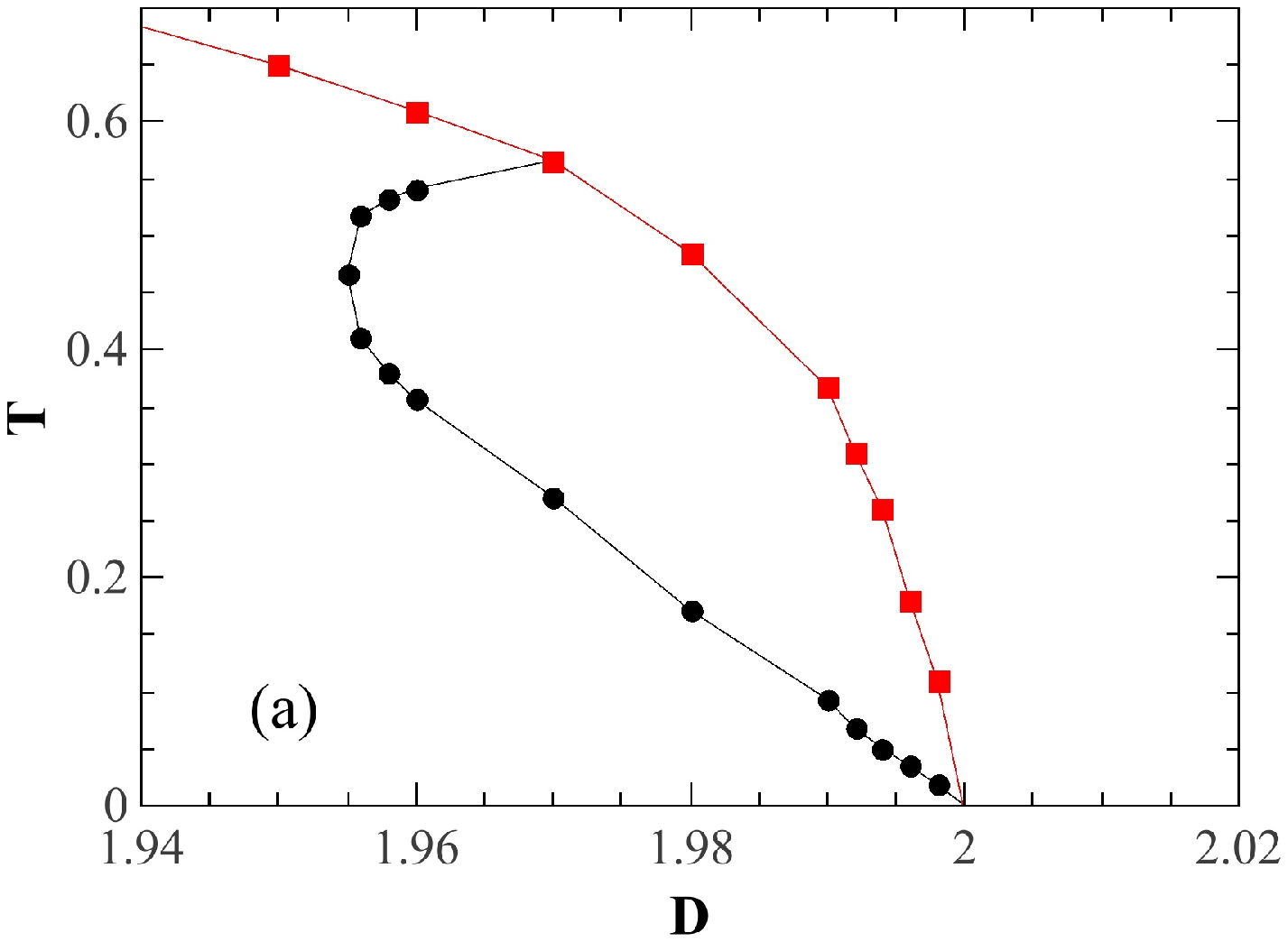}
\includegraphics[scale=0.4]{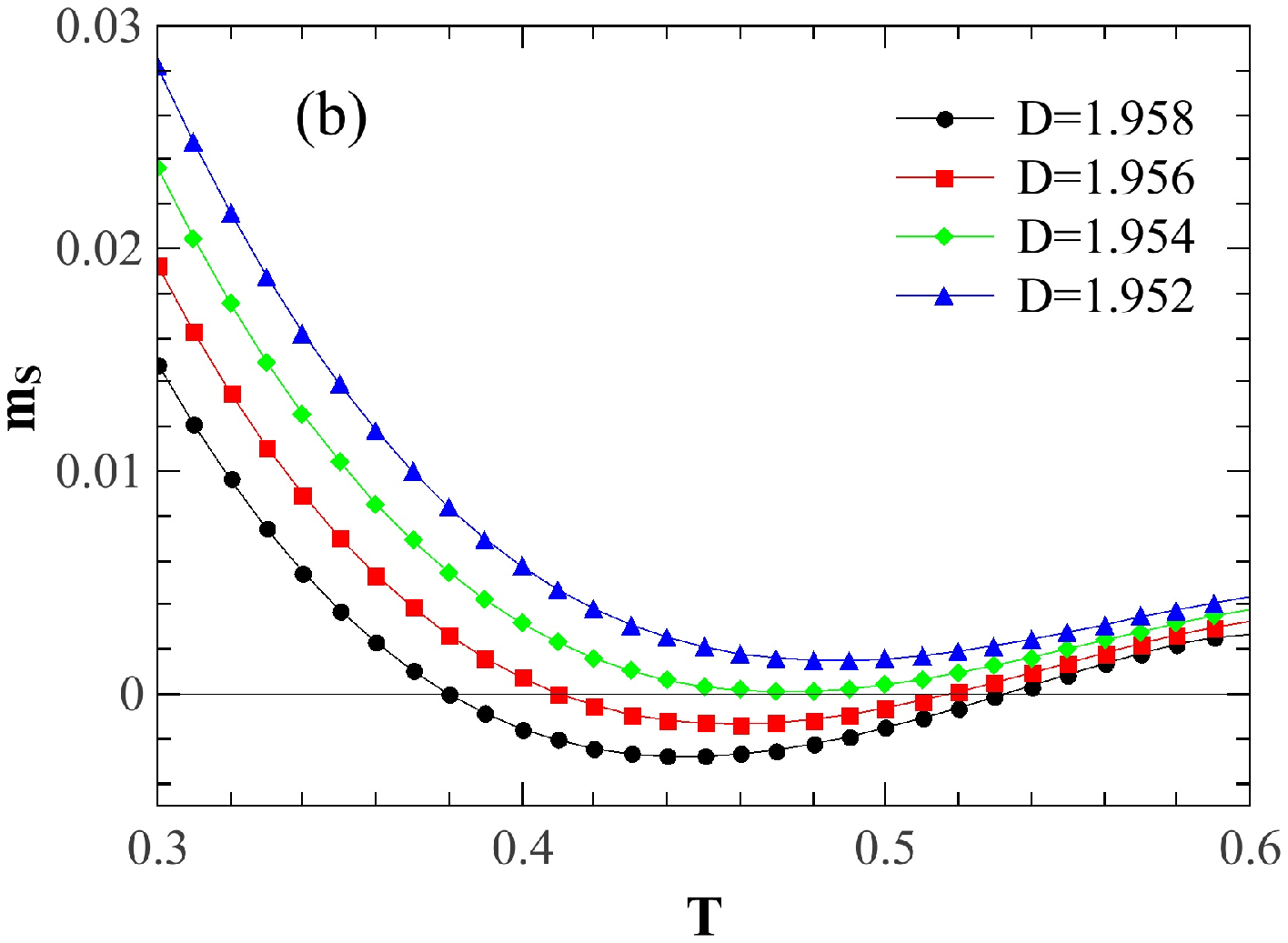}
\includegraphics[scale=0.4]{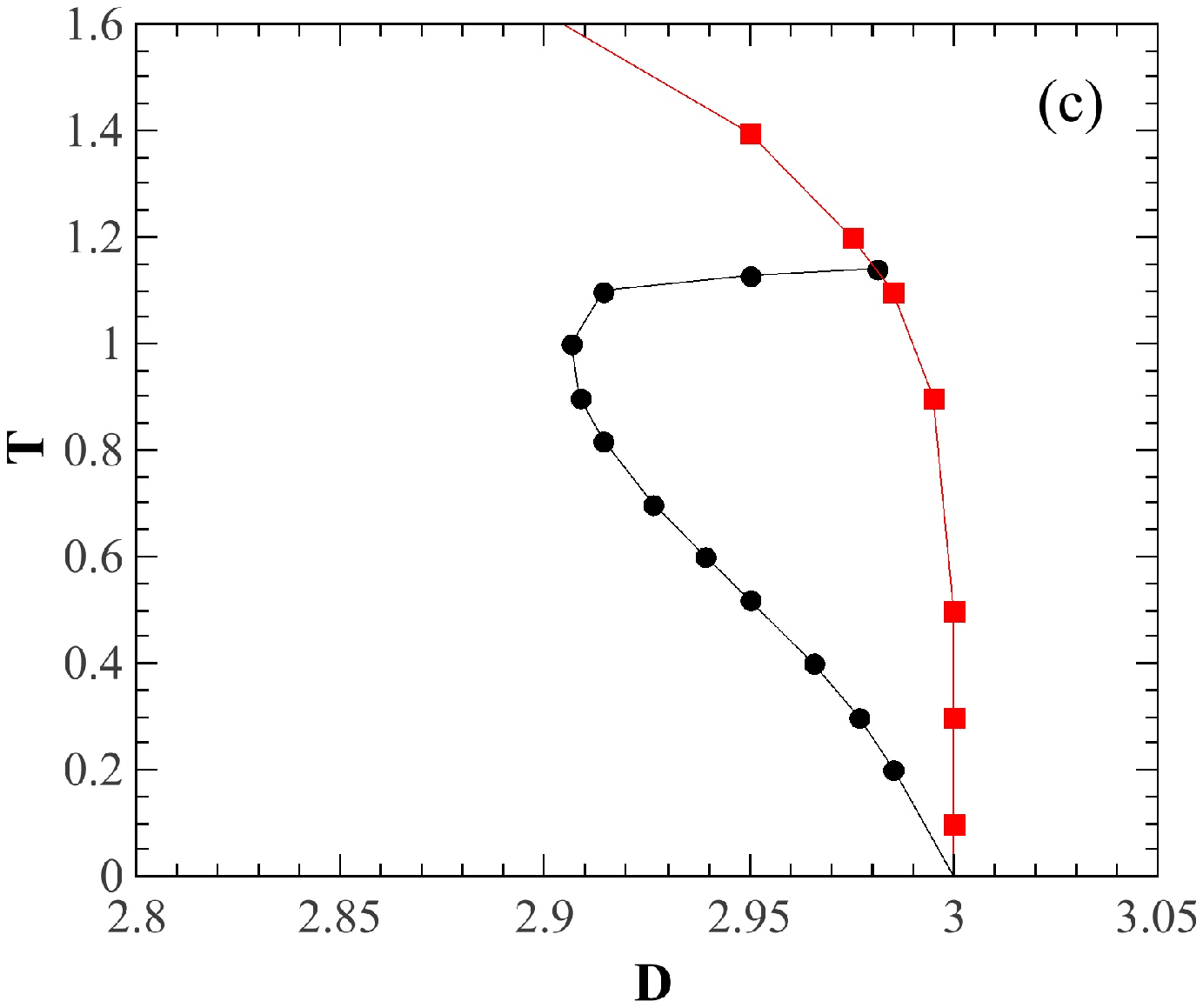}
\includegraphics[scale=0.4]{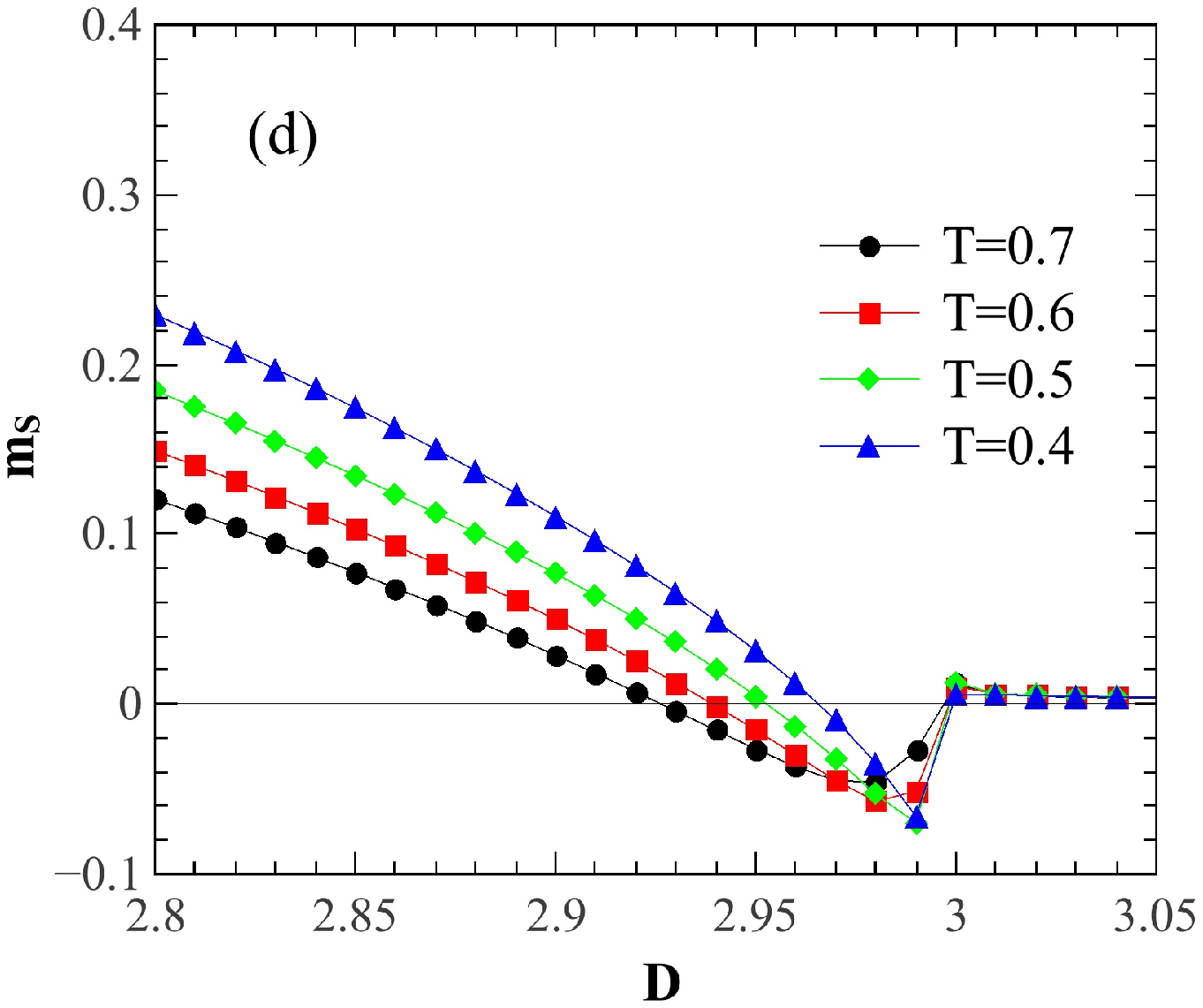}
\caption{(Color online) Temperature $T$ versus single-ion anisotropy strength $D$ on the square (a) and cubic (c) lattices. The square-solid lines denote the second-order transition lines and the circle-solid lines are compensation points. (b) Staggered magnetization $m_s$ versus temperature $T$ for different values of $D$, as shown in the figure. (d) Staggered magnetization $m_s$ versus single-ion anisotropy strength $D$ for different values of $T$, as shown in the figure. All results were obtained for $L=128$ (square lattice) and $L=32$ (cubic lattice).}
\label{fig4}
\end{figure}

We also have verified the existence of a compensation point as a function of the single-ion anisotropy strength $D$ on the square (Fig. \ref{fig4}(a)) and cubic (Fig. \ref{fig4}(c)) lattices. In Fig. \ref{fig4}(a) (square lattice), we can observe that there is no compensation point for $D < 1.954$. On the other hand, for values in the range of $1.954 < D \leq 1.970$ we have found always two compensation points and in the range of $1.970 < D < 2.0$ the system exhibits only one compensation point. To confirm, we have plotted the staggered magnetization $m_s$ versus temperature $T$, and for different values of $D$, where we can observe the two compensation points (see Fig. \ref{fig4}(b)). Now, for the case of the cubic lattice (see Fig. \ref{fig4}(c)) we  did not find any compensation point for $D < 2.9067$ whereas in the range of $ 2.9067 \leq D \leq 2.9147$ found two compensation points and in the range of $2.9147 <  D <  3.0$ only one compensation point. Here, we also have plotted the staggered magnetization $m_s$ versus single-ion anisotropy strength $D$, and for different values of $T$, as shown in Fig. \ref{fig4}(d). The system exhibits a compensation point.

\subsection{Phase diagram for the case $D_B$ fixed}

\begin{figure}[h]
\centering
\includegraphics[scale=0.56]{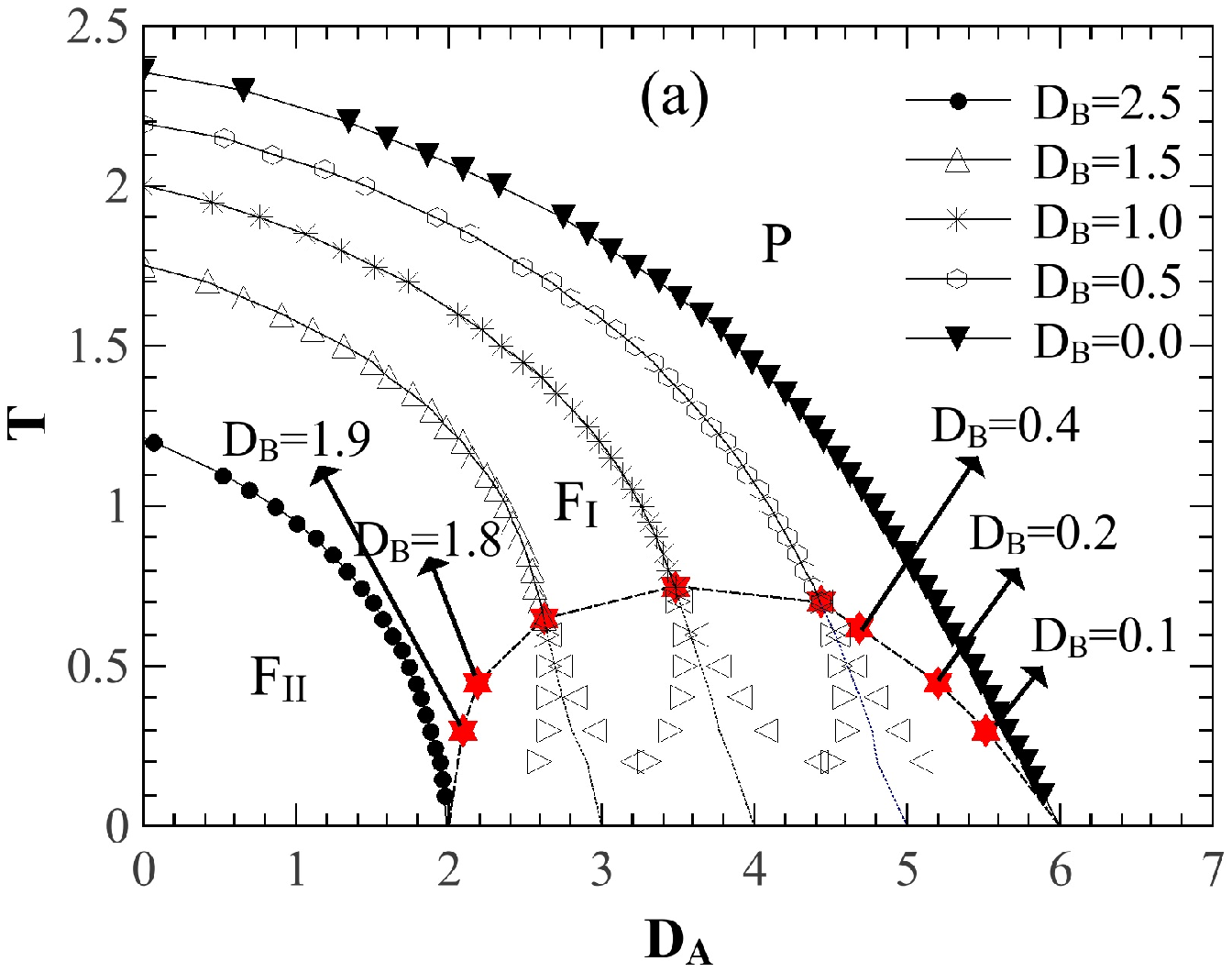}
\includegraphics[scale=0.56]{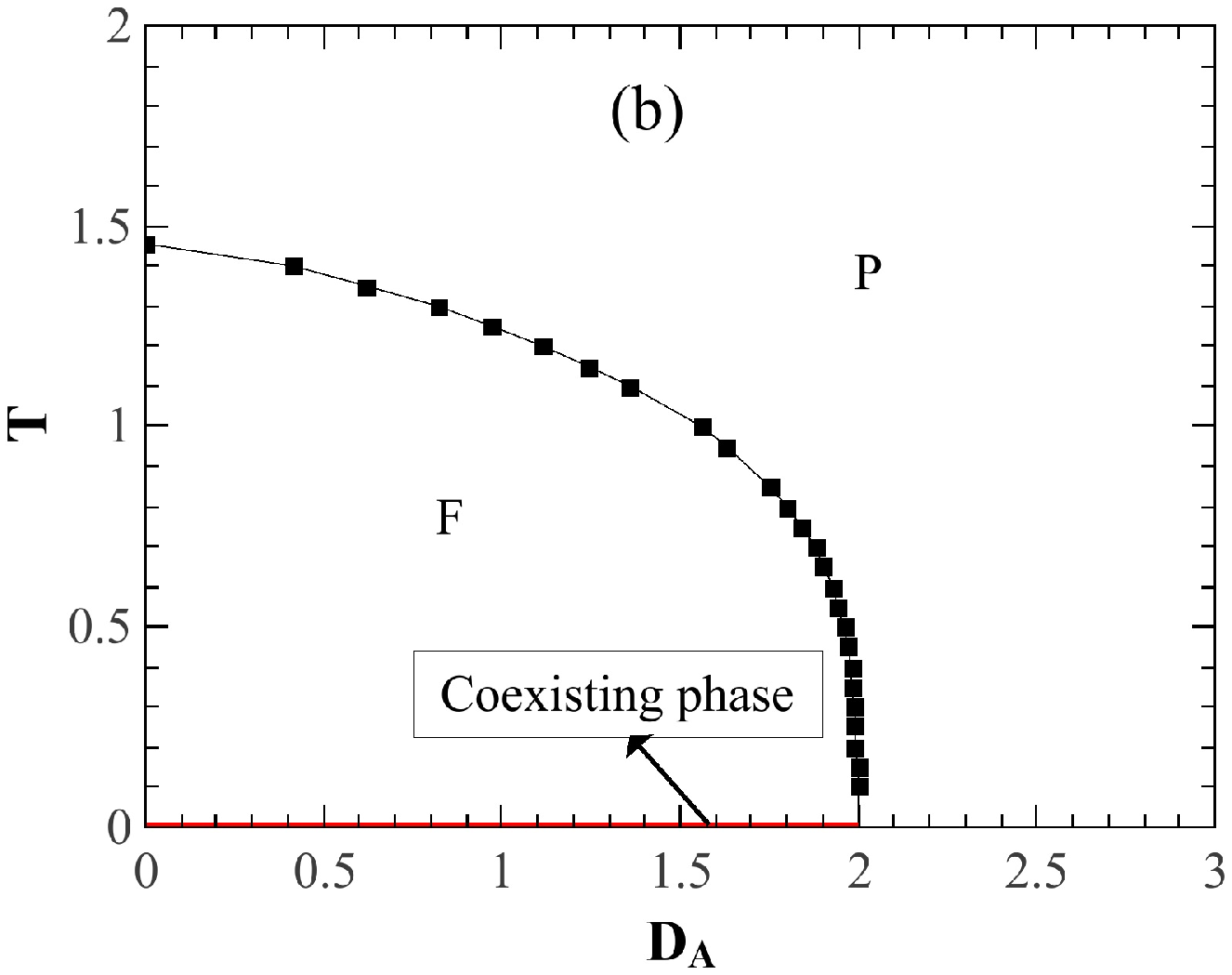}
\caption{(Color online) (a) Phase diagram in the $D_A-T$ plane for the mixed spin-1 and spin-3/2 Ising ferrimagnetic system on the square lattice, and for different values of $D_B$, as shown in the figure. All the full lines are second-order phase transitions. The empty triangles denote the hysteresis widths at first-order transitions between $F_I$ and $P$ phases with the expected phase transition boundary represented by the dotted lines. $F_I$, $F_{II}$ are the ordered phases and $P$ is a disordered phase. (b) Phase diagram for the special case $D_B =2.0$, where $F$ is an ordered phase and the solid line (for $T=0$) is the $F_I$ and $F_{II}$ coexisting phase.}
\label{fig5}
\end{figure}

Let us now consider the case in which $D_B$ is fixed. In Fig. \ref{fig5}(a), we have presented the phase diagram in the $D_A-T$ plane for the system only on the square lattice, and for different values of $D_B$ in the range of $0 < D_B < 2.50$. Therefore, in Fig. \ref{fig5}(a) is possible concluded that all phase transitions are second-order phase transitions between the ordered $F_I$ and disordered $P$ phases for $D_B \leq 0$, and between the ordered $F_{II}$ and disordered $P$ phases for $D_B \geq 2.5$. On the other hand, we have second- and first-order transitions between the $F_I$ and $P$ phases in the range of $ 0 < D_B < 2.0$.  In this region, the ordered $F_I$ phase is separated from the disordered $P$ phase by a line of phase transitions, which change at the tricritical point from second- to first-order. We calculated the coordinates of the tricritical points which are represented by star-dots, for example, the coordinates for some points are: with $D_B=1.50$ ($D^t_A=2.62$, $T^t=0.65$), $D_B=1.0$ ($D^t_A=3.48$, $T^t=0.75$) and $D_B=0.50$ ($D^t_A=4.43$, $T^t=0.70$). Thus, we can observe that there is a tricritical points line which is represented by a star-dotted line in the phase diagram. This line is in the range of $0.10 \leq D_B \leq 1.90$. In Fig. \ref{fig5}(b), we exhibited the phase diagram for the special case $D_B =2.0$ where we have defined $F$ (ferrimagnetic phase) as an ordered phase different from $F_I$ and $F_{II}$. The solid line (see $D_A$-axis for $T=0$) represents the coexistence from the $F_I$ and $F_{II}$ phases.

\begin{figure}[h]
\centering
\includegraphics[scale=0.38]{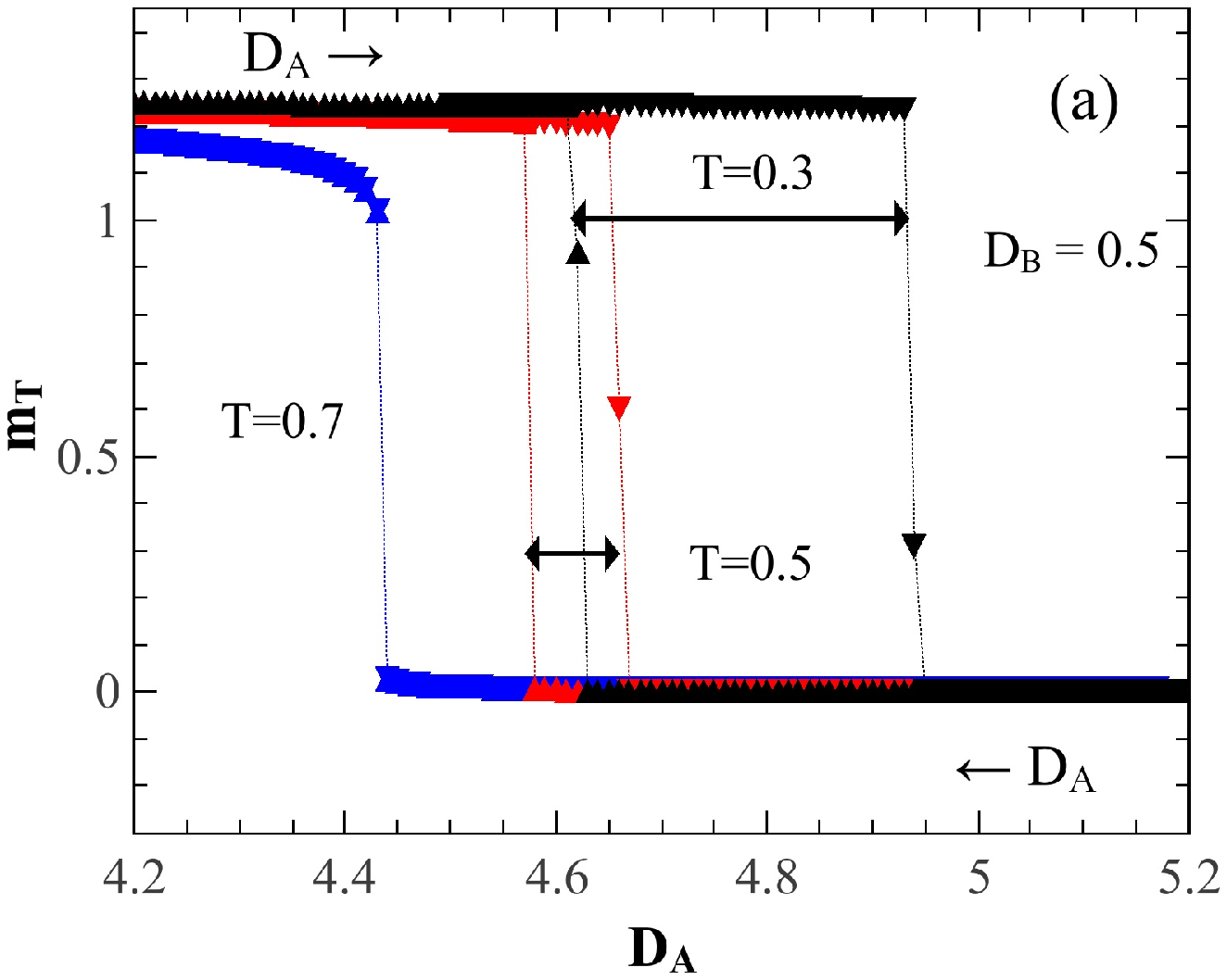}
\includegraphics[scale=0.38]{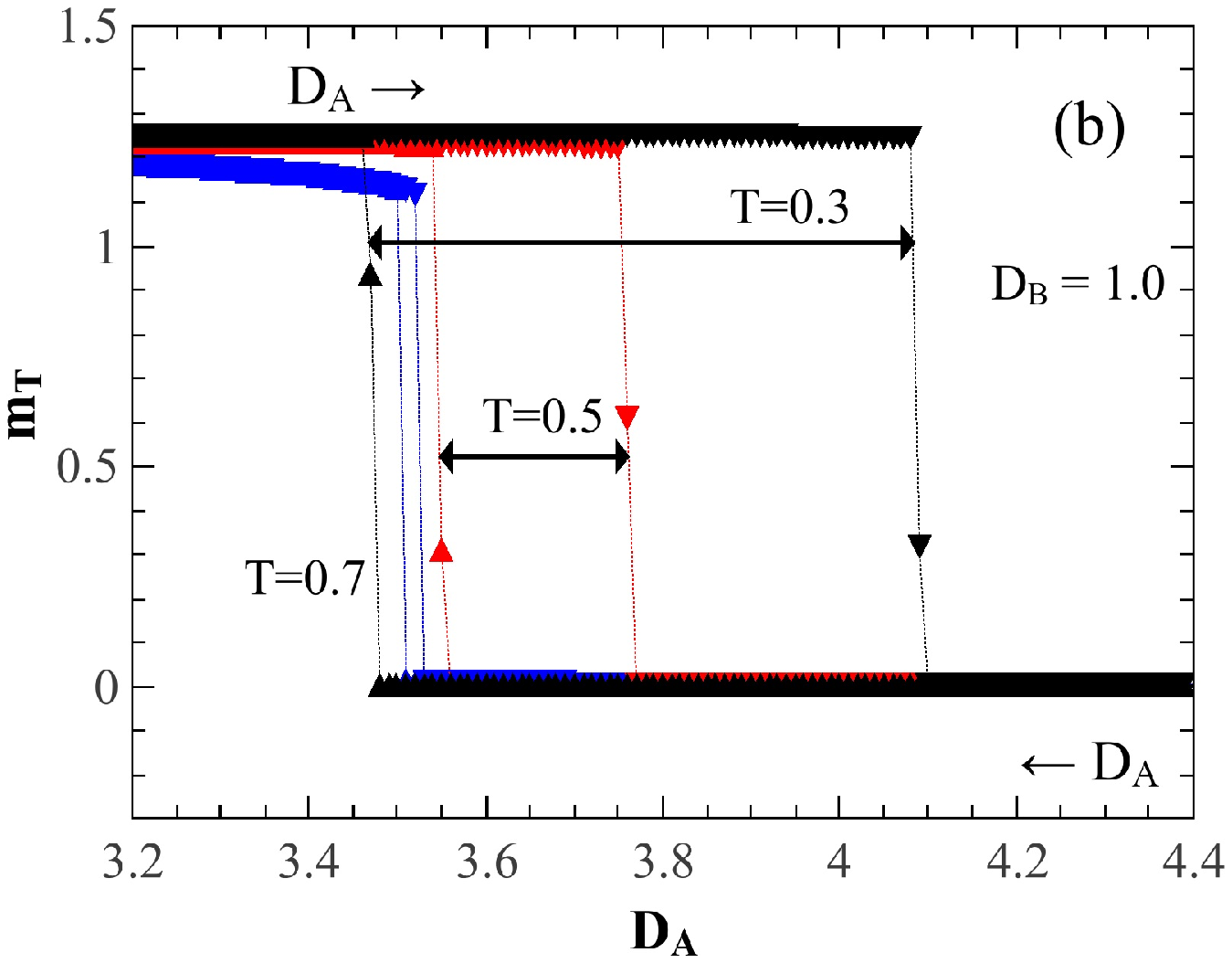}
\includegraphics[scale=0.38]{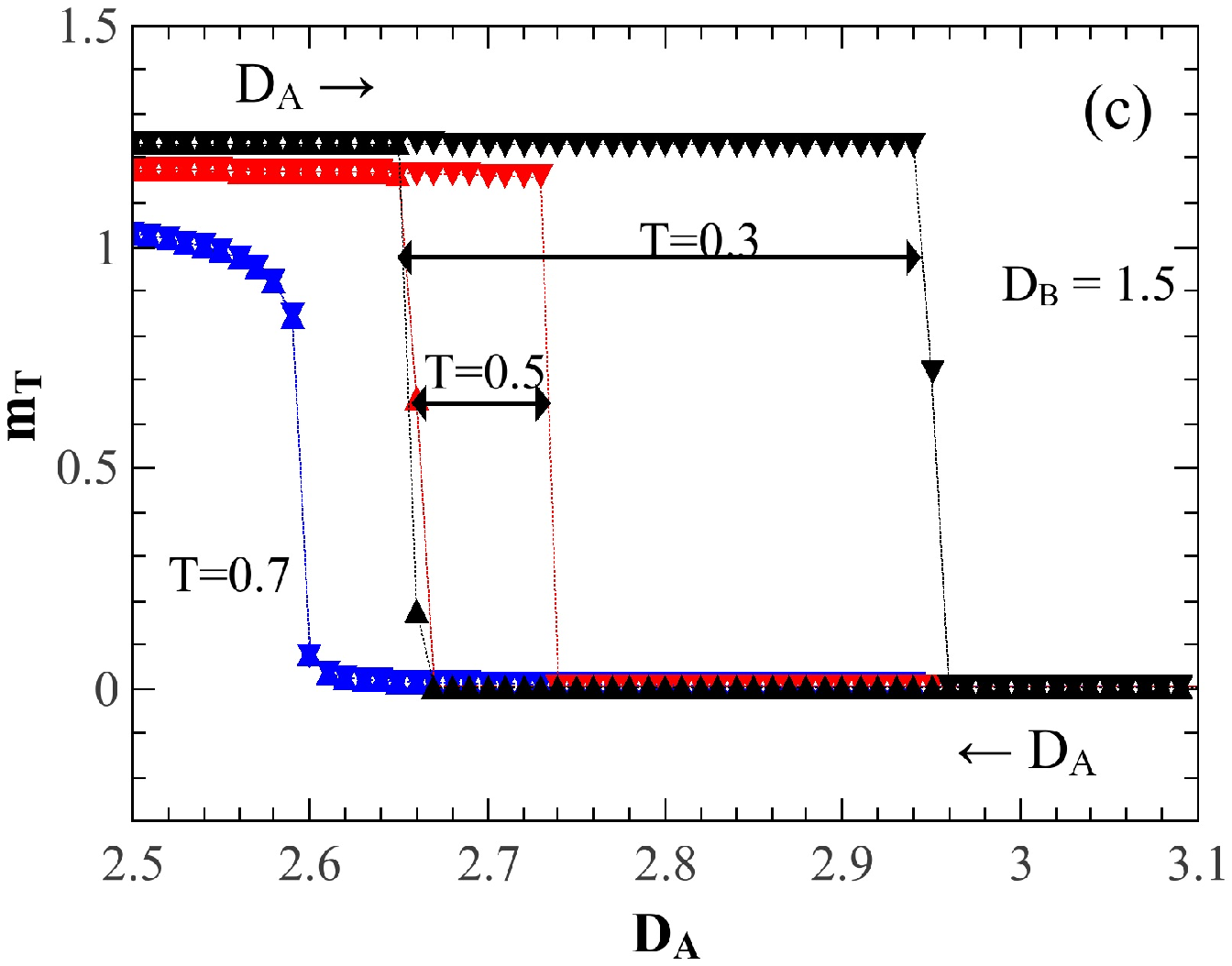}
\caption{(Color online) Hysteresis of the total magnetization $m_T$ as a function of increasing ($\bigtriangledown$) and decreasing ($\bigtriangleup$) single-ion anisotropy $D_A$ and for several values of $T$ fixed indicated in the figures. (a) It is for an anisotropy fixed $D_B=0.5$, (b) for 1.0 and (c) for 1.5. Here, we have used $L=128$ (square lattice) and the double-headed arrows denote the loop hysteresis widths. The dotted lines are a guide to the eyes.}
\label{fig6}
\end{figure}

Now looking at the transition between the two phases $F_I$ and $P$ for low-temperature, Fig. \ref{fig5}(a). We  presented in Fig. \ref{fig6} the total magnetization $m_T$ as increasing ($\bigtriangledown$) and decreasing ($\bigtriangleup$) functions of $D_A$ one can observe their discontinuous character and the appearance of hysteresis loops, the widths of which increase with decreasing temperature $T$. Therefore, an expected first-order transition boundary is obtained by a simple linear interpolation between the estimated tricritical and the exact ground-state transition points (square-dots) and only serves as a guide to the eye. We obtained these results for $D_B=0.50$ (Fig. \ref{fig6}(a)), 1.00 (Fig. \ref{fig6}(b)) and 1.50 (Fig. \ref{fig6}(c)). We have used $L=128$ (square lattice) and the double-headed arrows denote the loop hysteresis widths.

\subsection{Phase diagram for the case $D_A$ fixed}

\begin{figure}[h]
\centering
\includegraphics[scale=0.38]{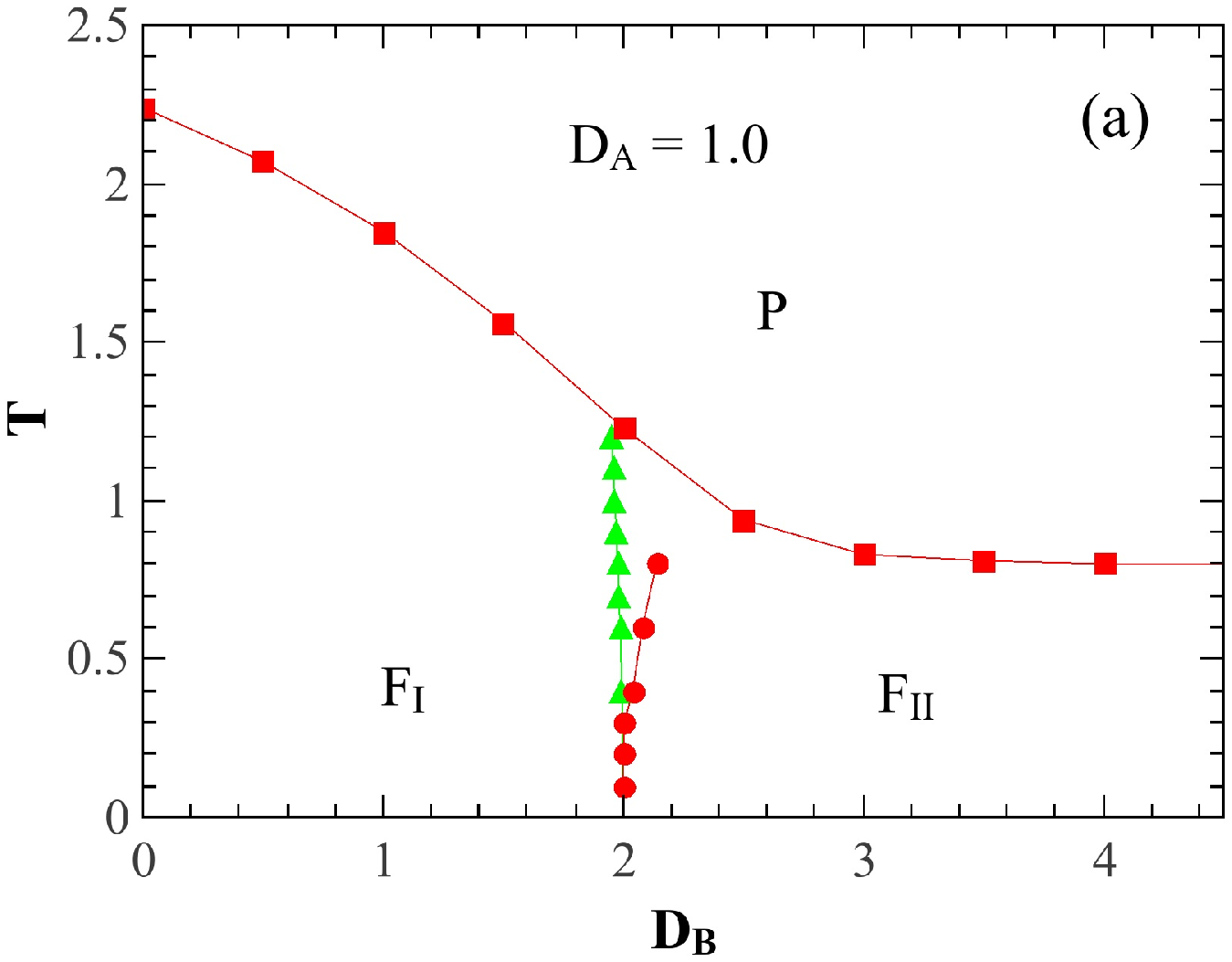}
\includegraphics[scale=0.38]{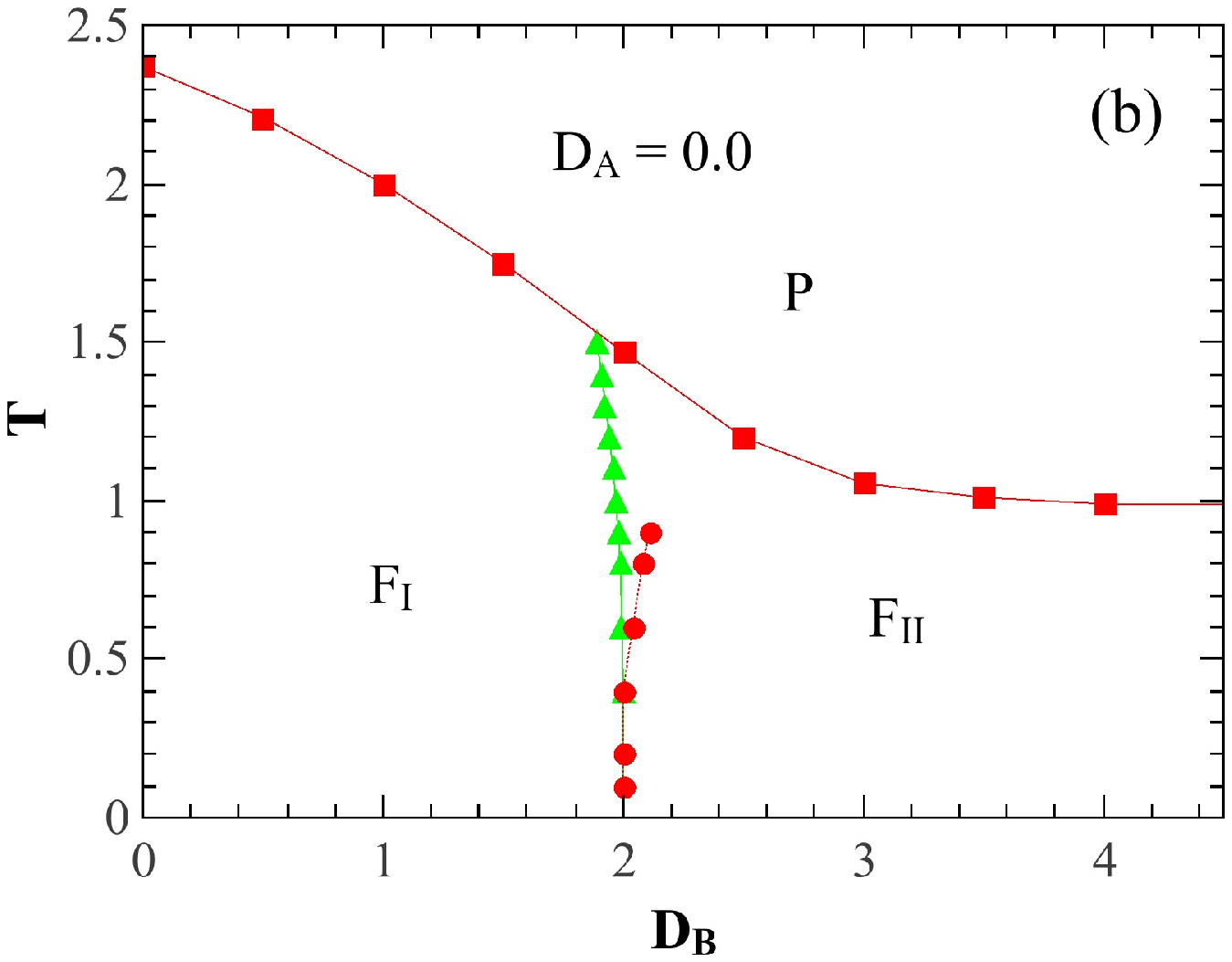}
\includegraphics[scale=0.38]{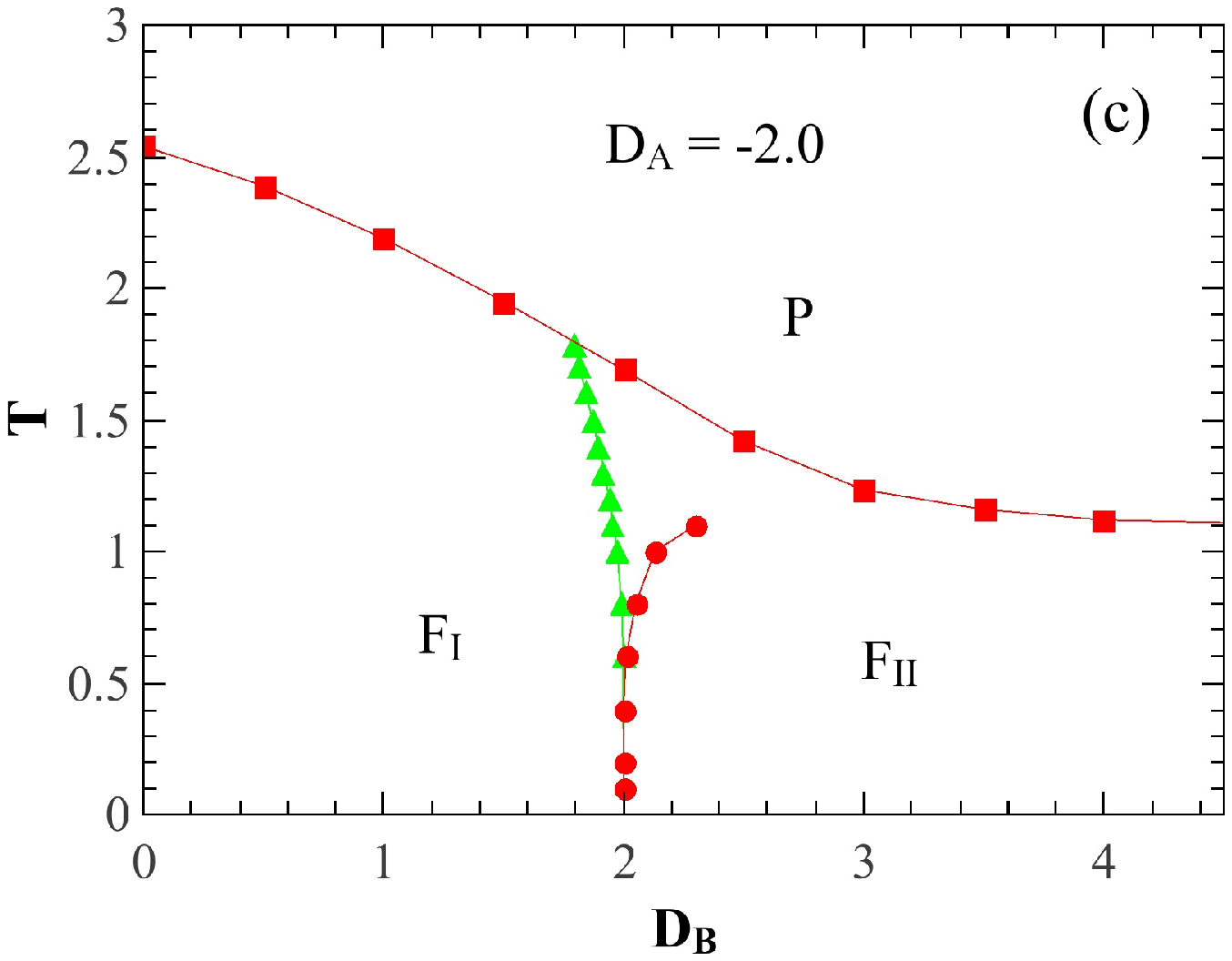}
\caption{(Color online) Phase diagram in the $D_B-T$ plane for the mixed spin-1 and spin-3/2 Ising ferrimagnetic system on the square lattice for different values of $D_A$, as shown in the figures: (a) for $D_A=1.00$, (b) 0 and (c) -2.00. All the square- and circle-solid lines are second-order phase transition. $F_I$ and $F_{II}$ are ordered phases and $P$ is a disordered phase. The triangle-solid lines are the compensation points.}
\label{fig7}
\end{figure}

Finally in this section, we exhibited the phase diagram in the $D_B-T$ plane for the system only on the square lattice and different values of $D_A$, as shown in Fig. \ref{fig7} and \ref{fig9}. Firstly, the phase diagrams of Fig. \ref{fig7}(a) with $D_A=1.0$, Fig. \ref{fig7}(b) with $D_A=0$ and Fig. \ref{fig7}(c) with $D_A=-2.0$ we found only second-order phase transitions from the ordered $F_I$ and $F_{II}$ to disordered $P$ phases, and between the ordered $F_I$ to $F_{II}$ phases. We also calculated the compensation points which are represented by triangle-solid lines. 

\begin{figure}[h]
\centering
\includegraphics[scale=0.5]{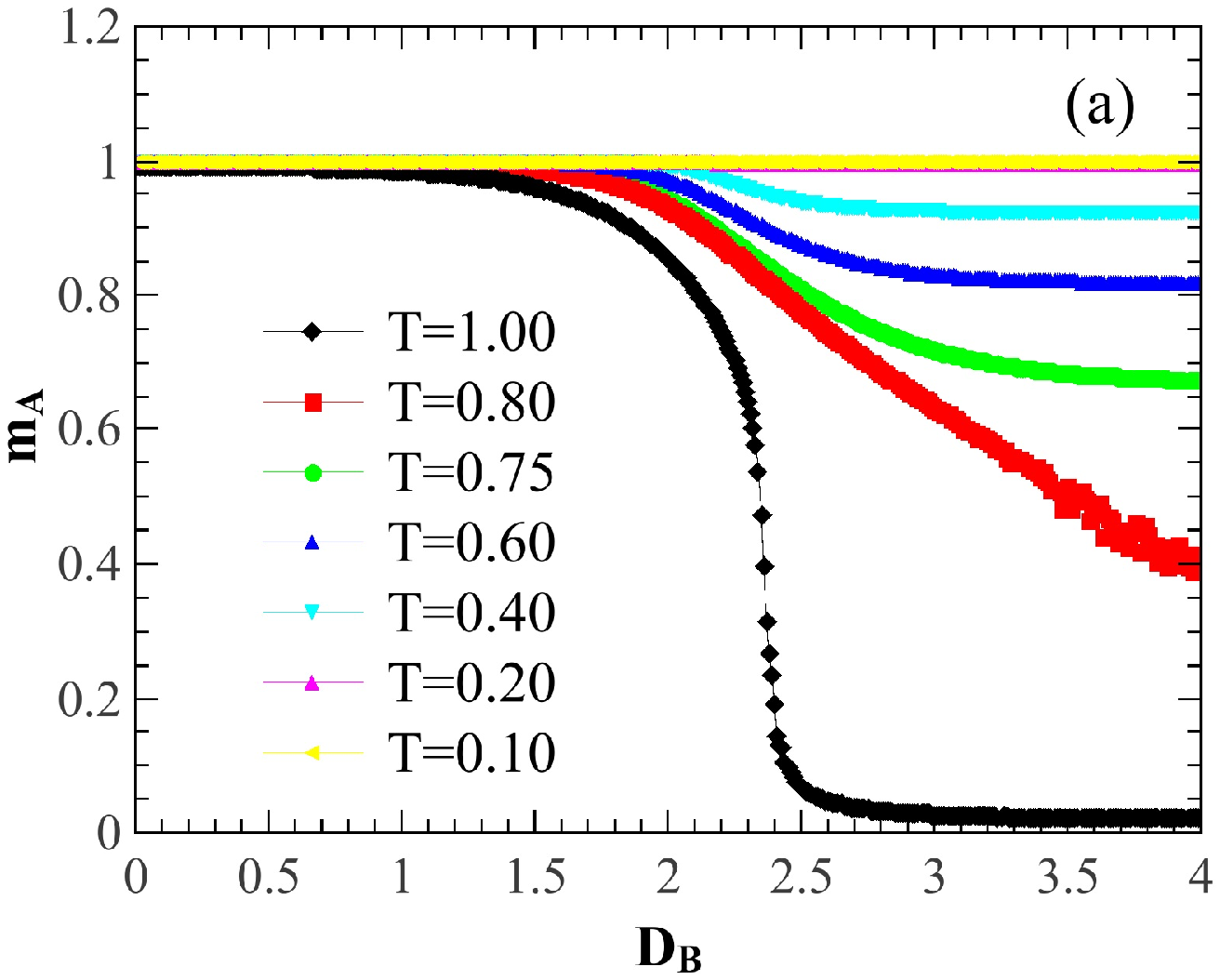} 
\includegraphics[scale=0.5]{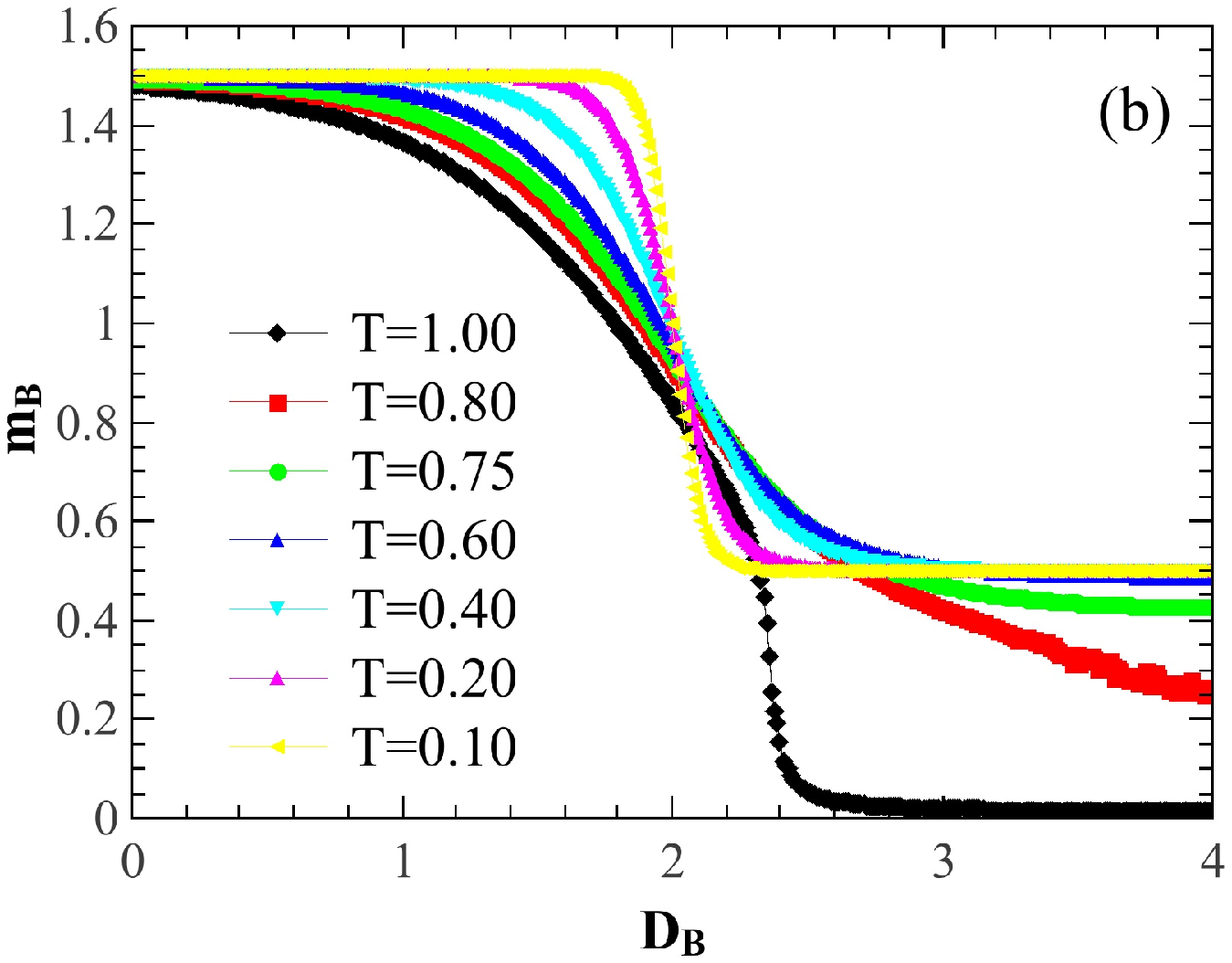}
\includegraphics[scale=0.5]{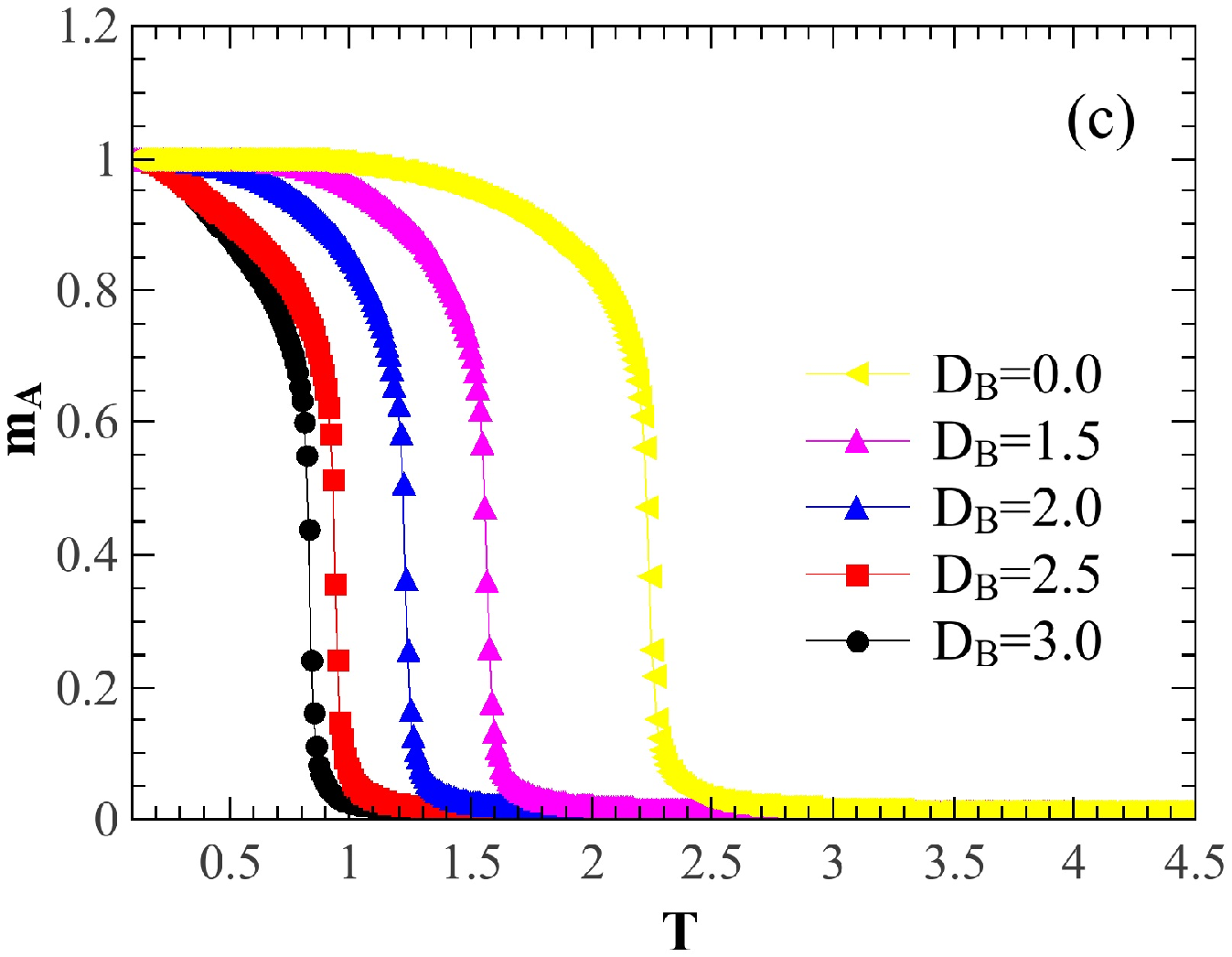}
\includegraphics[scale=0.5]{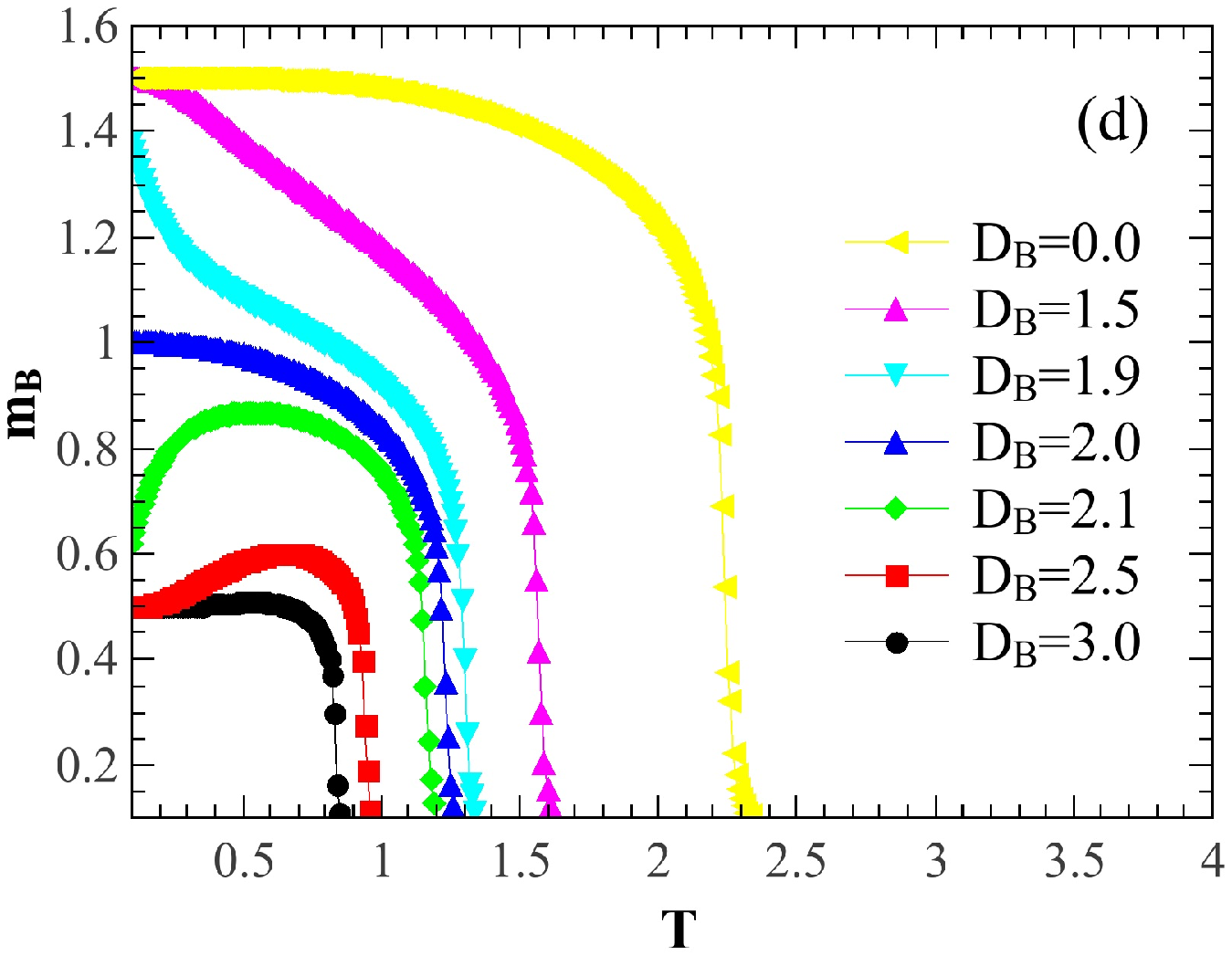}
\caption{(Color online) Sublattice magnetizations $m_A$ (a) and $m_B$ (b) as a function of the anisotropy $D_B$ for several values of $T$, as indicated in the figures. Sublattice magnetizations $m_A$ (c) and $m_B$ (d) as function of temperature $T$ for several values of $D_B$, as indicated in the figures. Here, we have used a lattice size $L = 128$ (square lattice) and fixed $D_A=1.0$.}
\label{fig8}
\end{figure}

It is important now to examine the behavior of the sublattice magnetizations $m_A$ and $m_B$ in the second-order phase transition regions and in order to verify the phase diagrams presented in Fig. \ref{fig7}. However, we have restricted our study to the behavior of the magnetization only for one value of the parameter $D_A=1.0$ (Fig. \ref{fig7}(a)). Thus, the behavior of $m_A$ and $m_B$, in these transitions, are shown in Fig. \ref{fig8}. In Fig. \ref{fig8}(a) and (b) we exhibited the behavior $m_A$ and $m_B$ as a function of the anisotropy $D_B$ for several values of $T$. In the range of  $0 \leq T \leq 0.2$ the sublattice magnetizations are constants $m_A = 1.0$ and $m_B = 1.5$ in the $F_I$ phase ($D_B \leq 2.0$, see Fig. \ref{fig7}). When the temperature increases $m_A = 1.0$ no change and $m_B$ changes quickly  for $m_B = 0.5$ now in the $F_{II}$ phase ($D_B >2.0$). This fact indicates that the system preserves the characteristics of the ground-state phases. On the other hand, in the range of  $0.2 < T \leq 0.8$, the sublattice magnetizations are $m_A \approx 1.0$ and $m_B \approx 1.5$ in the $F_I$ phase and they go to any value $0 <  m_A < 1.0$ and $0 < m_B < 0.5$ in the $F_{II}$ phase. This transition line, $F_I-F_{II}$ ends at the critical end point which the coordinate are: for $D_A=1.0$ ($D^e_B=2.14,T_e=0.80$), $D_A=0$ ($D^e_B=2.11,T_e=0.90$) and $D_A=-2.0$ ($D^e_B=2.30,T_e=1.10$). Now, for $T > 0.9$ the $m_A$ and $m_B$ start in the $F_I$ phase, cross to the $F_{II}$ phase and after go to $P$ phase, as for example, we can see in the curve for $T=1.0$, Figs.~\ref{fig8}(a) and (b). 

We also can observe the behavior sublattice magnetizations in the transition line $F_I$, $F_{II}$ to $P$. The temperature dependence of the $m_A$ and $m_B$ and for several values of $D_B$ are shown in Figs. \ref{fig8}(c) and (d), respectively. For $T<T_c$ the $m_B$ has a re-entrant behavior, as can see for $D_B=2.1$ and $D_B=2.5$ in Fig. \ref{fig8}(d). For this case, we have a competition between the parameters $J$, $D_A$, $D_B$, and $T$ which makes it difficult to align the sublattice magnetization and also due to the finite-size effects in low-temperature. When the temperature is increased ($T \rightarrow T_c$), the system passes from the $F_{II}$ phase to the $F_I$ phase and goes to zero in the $P$ phase.
 
\begin{figure}[h]
\centering
\includegraphics[scale=0.7]{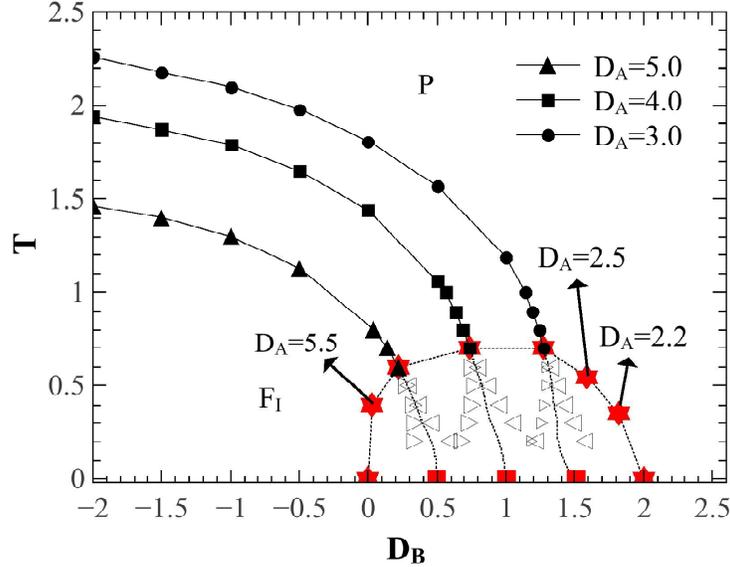} 
\caption{(Color online) Phase diagram in the $D_B-T$ plane for the mixed spin-1 and spin-3/2 Ising ferrimagnetic system on the square lattice, and different values of $D_A$ indicated in the figure. All the full lines are second-order phase transitions. The empty triangles denote the hysteresis widths at first-order transitions between $F_I$ and $P$ phases with the expected phase transition boundary represented by the dotted lines. The star-dotted line represents a tricritical points line. The  $F_I$ and $P$ are the ordered and disordered phases, respectively.}
\label{fig9}
\end{figure}

Another region of the phase diagram which also is important, but it was not presented in Fig.~\ref{fig7}, is shown in Fig.~\ref{fig9} for values of  $D_A \geq 2.0$ as indicated in the figure. For values of $D_B$ between 0 and 2.0, we also have second- and first-order phase transition lines linked by a tricritical point and they separate the $F_I$ phase from the $P$ phase. We have connected these points by a line (star-dotted line) and showed that there is a small region where all the phase transition lines are first-order. We also can observe in the phase diagram that the empty triangles represent the hysteresis loop widths at first-order transitions between $F_I$ and $P$ phases. This transition line was obtained using the same procedure as in the previous section and the expected phase transition boundary is denoted by a dotted line. 

\begin{figure}[h]
\centering
\includegraphics[scale=0.38]{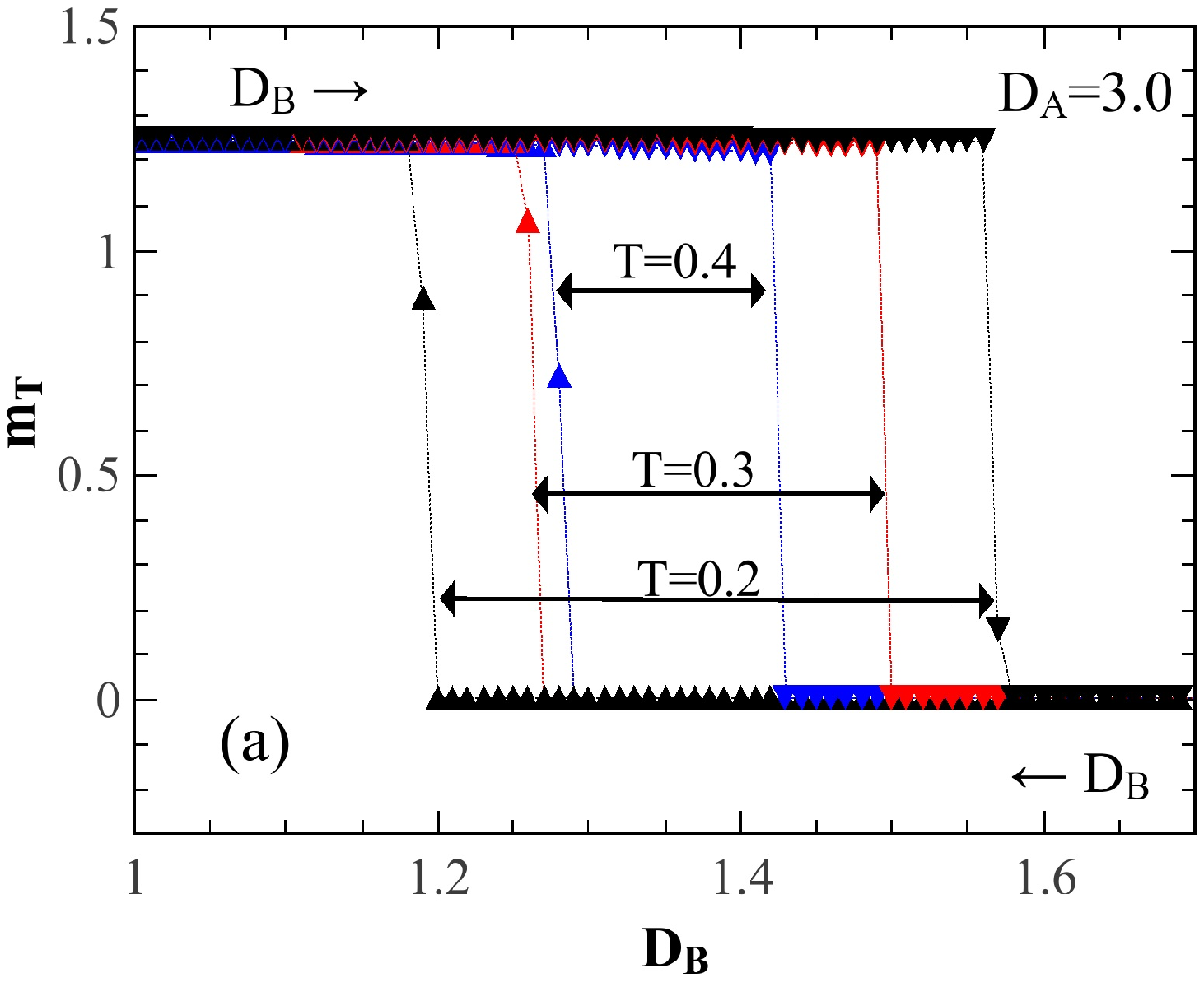}
\includegraphics[scale=0.38]{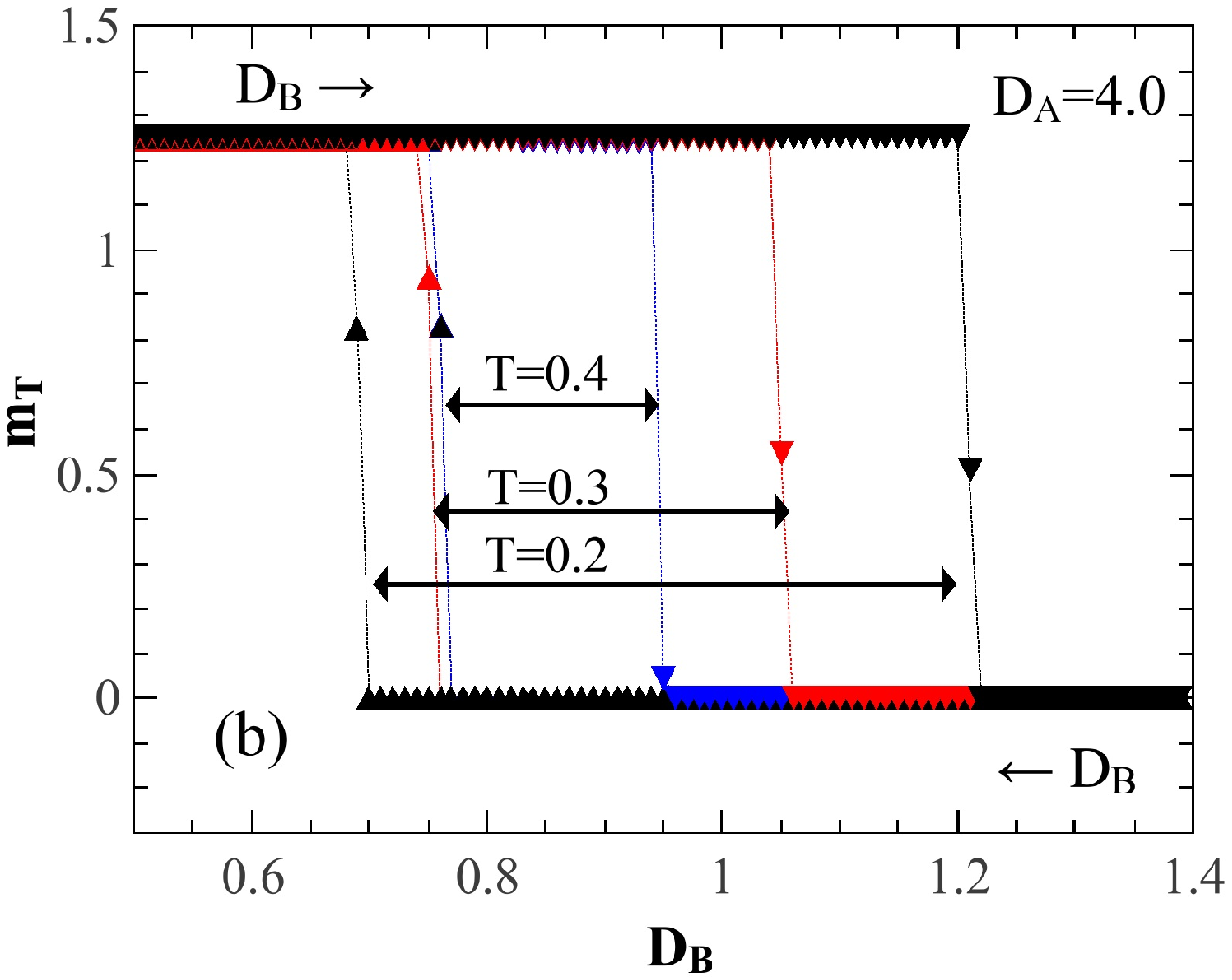}
\includegraphics[scale=0.38]{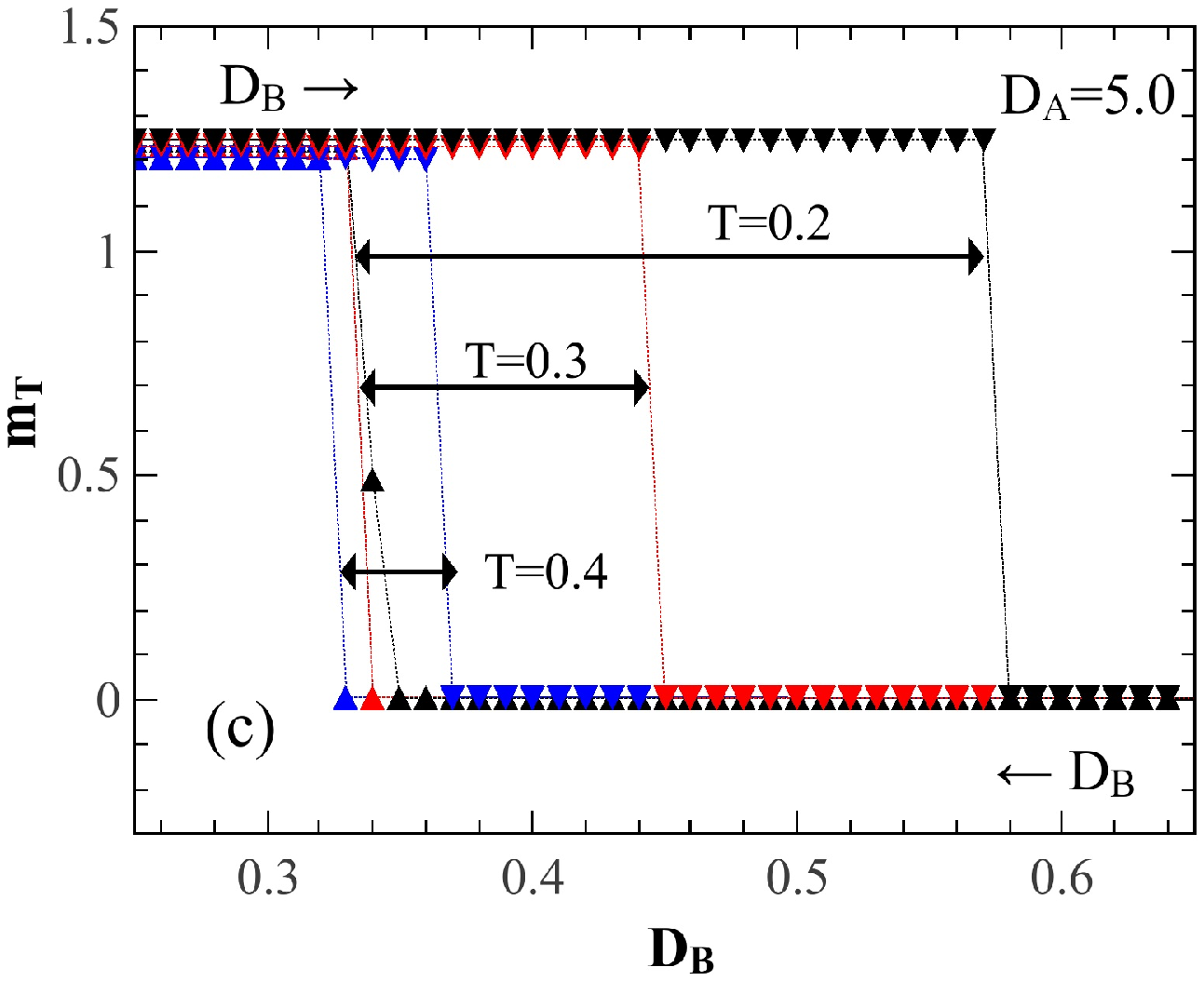}
\caption{(Color online) Hysteresis loop of the total magnetization $m_T$ as a function of increasing ($\bigtriangledown$) and decreasing ($\bigtriangleup$) single-ion anisotropy $D_B$ and for several values of temperatures $T$ fixed indicated in the figures. (a) It is for an anisotropy fixed $D_A=3.0$, (b) for 4.0 and (c) for 5.0. Here, we have used $L=128$ (square lattice) and the double-headed arrows denote the loop hysteresis widths. The dotted lines are a guide to the eyes.}
\label{fig10}
\end{figure}

In Fig~\ref{fig10}, we exhibited the hysteresis loop of the total magnetization $m_T$ as a function of increasing ($\bigtriangledown$) and decreasing ($\bigtriangleup$) single-ion anisotropy $D_B$ and for several values of temperatures $T$ fixed indicated in the figures. Fig. \ref{fig10}(a) is for an anisotropy fixed $D_A=3.0$, Fig. \ref{fig10}(b) for $D_A=4.0$ and the Fig. \ref{fig10}(c) for $D_A=5.0$. These plots show a characteristic behaviors in first-order transitons, such as the discontinuities in the magnetizations. 

\section{Conclusions}

In this work, we have studied the mixed-spin Ising model with ferrimagnetic interaction between spin-1 (states $\pm 1$, 0) and spins-3/2 (states $\pm 1/2 $, $\pm 3/2$). We performed Monte Carlo simulations on the square and cubic lattices, where each type of spin is fixed in a sublattice and with anisotropies $D_A$ and $D_B$ on the respective sublattices A and B. Firstly, we studied a particular case of model in which the anisotropies are equal, $D = D_A/J = D_B/J$. This case was studied by Zukovic {\it et al.} \cite{milo} on the square lattice and they showed that the system presents only second-order phase transition. We also found only second-order phase transitions for both square and cubic lattices between the ordered $F$ and disordered $P$ phases, and a multi-compensation point behavior, i.e., with two compensation points for the same value of anisotropy $D$ (see Fig. \ref{fig4}(a) and (c)). In the case of anisotropy $D_B$ fixed, the phase diagram presents two different ferrimagnetic phases, $F_I$ and $F_{II}$. In the range of $0 < D_B < 2.0$, we found first- and second-order phase transitions between the ordered $F_I$ and the disordered $P$ phases, i.e., the system presents a tricritical behavior. On the other hand, in the range of $- \infty \leq D_B \leq 0$ and $2.0 < D_B < \infty$ we have found only second-order phase transitions between the  $F_I - P$ and $F_{II}-P$ phases, respectively (see Fig. \ref{fig5}). We also observe that in this case, the model does not exhibit compensation points.

Now, for the case of fixed $D_A$, but in the range of $- \infty < D_A < 2.0$ (see Fig. \ref{fig7}), we observe second-order phase transitions between the ordered $F_I$ and $F_ {II}$ phases for low-temperatures $T \leq 0.3$, where the system preserves the characteristics of the ground-state phases ($T = 0$). When we increase the temperature $T > 0.3$ (the system no more preserves the characteristic of the ground state). The system continuous exhibiting second-order phase transitions between the $F_I - F_{II}$ phases (see Fig. \ref{fig8}) where the phase $F_I$ is defined as a region with $m_A \neq 0$ ($0 < m_A  < 1.0$) and $m_B \neq 0$ ($ 0 < m_B < 1.5 $) and $F_{II}$ a region with $m_A \neq 0$ ($0 < m_A < 1.0$) and $m_B \neq 0$ ($0 < m_B < 0.5$) for different temperature values $T$ and for any value of $D_B > 2.0$. We also found second-order phase transitions between the phases: $F_I-P$ and $F_{II}-P$. We also observed first- and second-order phase transitions between phase $F_I$ and phase $P$, in the range of $2.0 < D_A  < 6.0$  (see Fig. \ref{fig9}), i.e., it presents a tricritical behavior with a tricritical points line.

\section{Acknowledgments}

The authors acknowledge financial support from the Brazilian agencies CNPq and CAPES.

\end{document}